\documentclass[11pt]{article}

\usepackage{amsmath,amssymb,amsthm,bm}
\usepackage{mathtools}
\usepackage{graphicx}
\usepackage[font=small,labelfont=bf,width=\textwidth]{caption}
\usepackage{subcaption}
\usepackage{fullpage}
\usepackage{color}
\usepackage[colorlinks=true, citecolor=blue]{hyperref}
\usepackage[usenames,dvipsnames]{xcolor}
\usepackage{tikz}
\usetikzlibrary{calc}
\usetikzlibrary{arrows}
\usetikzlibrary{patterns}
\usetikzlibrary{decorations.pathmorphing}

\RequirePackage[sc]{mathpazo}
\RequirePackage[T1]{fontenc}

\newcommand{\CGf}{\ensuremath{\mathcal{C}}}

\newcommand{\ave}[1]{\langle #1 \rangle}

\newcommand{\ew}{z}
\newcommand{\cw}{u}
\newcommand{\cwc}{\cw_\mathrm{c}}
\newcommand{\ewc}{\ew_\mathrm{c}}

\newcommand{\ee}{\text{E}}
\newcommand{\ec}{\text{C}}
\newcommand{\ef}{\text{F}}
\newcommand{\el}{\text{L}}
\newcommand{\eu}{\text{U}}

\newcommand{\Rgs}{R^2_\mathrm{g}}
\newcommand{\Res}{R^2_\mathrm{e}}
\newcommand{\Rms}{R^2_\mathrm{m}}
\newcommand{\Mg}{M_\mathrm{g}}
\newcommand{\Me}{M_\mathrm{e}}
\newcommand{\Mm}{M_\mathrm{m}}

\newcommand{\bO}{\bm \Omega}
\DeclareMathOperator{\littleo}{o}
\DeclareMathOperator{\bigO}{O}

\newcommand\blfootnote[1]{%
  \begingroup
  \renewcommand\thefootnote{}\footnote{#1}%
  \addtocounter{footnote}{-1}%
  \endgroup
}

\usepackage{authblk}

\author{Nicholas R. Beaton\thanks{nrbeaton@unimelb.edu.au, ORCID:0000-0001-8220-3917} }
\author{Anthony J. Guttmann\thanks{guttmann@unimelb.edu.au, ORCID:0000-0003-2209-7192}}
\affil{School of Mathematics and Statistics, The University of Melbourne, Australia}

\author{Iwan Jensen\thanks{iwan.jensen@flinders.edu.au, ORCID:0000-0001-6618-8470}}
\affil{College of Science and Engineering, Flinders University, Australia}

\title{Two-dimensional interacting self-avoiding walks: new estimates for critical temperatures and exponents}

\begin{document}
\maketitle

\abstract{We investigate, by series methods, the behaviour of interacting self-avoiding walks (ISAWs) on the honeycomb lattice and on the square lattice. This is the first such investigation of ISAWs on the honeycomb lattice. We have generated data for ISAWs up to 75 steps on this lattice, and 55 steps on the square  lattice. For the hexagonal lattice we find the $\theta$-point to be at $\cwc = 2.767 \pm 0.002.$ The honeycomb lattice is unique among the regular two-dimensional lattices in that the exact growth constant is known for non-interacting walks, and is $\sqrt{2+\sqrt{2}}$ \cite{DS12}, while for half-plane walks interacting with a surface, the critical fugacity, again for the honeycomb lattice, is $1+\sqrt{2}$ \cite{BBDDG}. We could not help but notice that $\sqrt{2+4\sqrt{2}} = 2.767\ldots .$ We discuss the difficulties of trying to prove, or disprove, this possibility. 

For square lattice ISAWs we find $\cwc=1.9474 \pm 0.001,$ which is consistent with the best Monte Carlo analysis.

We also study bridges and terminally-attached walks (TAWs) on the square lattice at the $\theta$-point. We estimate the exponents to be $\gamma_b=0.00 \pm 0.03,$ and $\gamma_1=0.55 \pm 0.03$ respectively. The latter result is consistent with the prediction \cite{DS87, SS88, SSV93} $\gamma_1(\theta) = \nu =\frac47$, albeit for a modified version of the problem, while the former estimate is predicted in \cite{DG19b} to be zero.}

\section{Introduction}\label{sec:intro}

An\blfootnote{All series data generated for this paper are available with the preprint at \href{https://arxiv.org/abs/1911.05852}{arXiv:1911.05852}.} $n$-step self-avoiding walk (SAW) ${\bm \omega}$ on a regular lattice is 
a sequence of {\em distinct} vertices $\omega_0, \omega_1,\ldots , \omega_n$ 
such that each vertex is a nearest neighbour of its predecessor. 
SAWs are considered distinct up to translations of the starting point $\omega_0$.
We shall use the symbol ${\bm \Omega}_n$ to mean the set of all 
SAWs of length $n$, and denote $c_n = |\bO_n|$.

It is well known that the (lattice-dependent) connective constant $\mu = \lim_{n\to\infty}(c_n)^{\frac{1}{n}}$ exists, and for the honeycomb lattice $\mu=\sqrt{2+\sqrt{2}}$~\cite{DS12}. It is unproved but widely accepted that the sub-dominant asymptotic behaviour of $c_n$ is governed by a power law, namely
\begin{equation}
    c_n = Cn^{\gamma-1}\mu^n(1+\littleo(1))
\end{equation}
where the constant $C$ is lattice-dependent but the exponent $\gamma$ is universal, and depends only on the dimension.\footnote{Note that we will sometimes write $c_n \sim Cn^{\gamma-1}\mu^n$; more generally, for two sequences $a_n$ and $b_n$ we write $a_n \sim b_n$ if $a_n/b_n \to 1$ as $n\to\infty$.} Without loss of generality we will orient our honeycomb lattice so that it has vertical edges, and assume that SAWs start at a vertex at the bottom of a vertical edge (see Figure~\ref{fig:hcisaw}). We also scale the lattice so that edges have unit length.

We are also interested in two subsets of $\bO_n$: \emph{terminally attached walks} (TAWs) and \emph{bridges}. For the purposes of defining these objects it is easier to transform the honeycomb lattice into a \emph{brickwork} lattice (see Figure~\ref{fig:hcisaw}), with walks starting at a vertex with edges in the $\pm x$ and $+y$ directions. If $\omega$ is a SAW on the honeycomb lattice then let $\omega'$ be its transformation to the brickwork lattice, and denote $\omega'_i = (x'_i,y'_i)$. A TAW is then a SAW $\omega$ such that $y'_i \geq y'_0$ for all $i=0,1,\dots,n$, while a bridge satisfies the stronger condition $y'_0 < y'_i \leq y'_n$ for all $i=1,\dots,n$. If we let $t_n$ be the number of TAWs of length $n$ and $b_n$ be the number of bridges, then it is known that $\lim_{n\to\infty}(t_n)^{\frac1n} = \lim_{n\to\infty}(b_n)^{\frac1n} = \mu$, and expected that
\begin{equation}
    t_n = Tn^{\gamma_1-1}\mu^n(1+\littleo(1)) \qquad\qquad b_n = Bn^{\gamma_b-1}\mu^n(1+\littleo(1))
\end{equation}
where again $T$ and $B$ depend on the lattice while $\gamma_1$ and $\gamma_b$ depend only on the dimension. These conjectures are based on Coulomb gas and SLE formulations.

While the existence of the aforementioned critical exponents has not been proved, their exact values in two dimensions have been conjectured. Indeed, it is expected that $\gamma=\frac{43}{32}$ \cite{Nienhuis1982}, $\gamma_1=\frac{61}{64}$ \cite{Cardy1984}, and $\gamma_b = \frac{9}{16}$ \cite{BGJL}. Further support for the quoted value of $\gamma_b$ is given in \cite{DG19}.

\subsection{Geometric properties}

We will be interested in the geometric properties of SAWs and ISAWs, and in particular three measurements of `size': the (squared) radius of gyration, end-to-end distance, and monomer-to-end distance. For a given SAW $\omega$ of length $n$ these are given by
\begin{align}
    \Rgs(\omega) &= \frac{1}{2(n+1)^2}\sum_{i,j=0}^n|\omega_i-\omega_j|^2 \\
    \Res(\omega) &= |\omega_0-\omega_n|^2 \\
    \Rms(\omega) &= \frac{1}{2(n+1)}\sum_{i=0}^n\left(|\omega_0-\omega_i|^2 - |\omega_n-\omega_i|^2\right)
\end{align}
where $|\cdot|$ denotes Euclidean distance (we scale the lattice so that edges have unit length). The mean values of these quantities are then naturally
\begin{align}\label{eqn:mean_metric_quantities}
    \ave{\Rgs}_n = \frac{1}{c_n} \sum_{\omega\in\bO_n}\Rgs(\omega), & &
    \ave{\Res}_n = \frac{1}{c_n} \sum_{\omega\in\bO_n}\Res(\omega), & &
    \ave{\Rms}_n = \frac{1}{c_n} \sum_{\omega\in\bO_n}\Rms(\omega).
\end{align}
Each is expected to display power-law behaviour, ie.~$\ave{R^2_x}_n \sim M_x n^{2\nu}$ (where $x$ is one of g, e, or m) where $M_x$ depends on the lattice, while $\nu$ depends only on dimension. In two dimensions it is expected that $\nu=\frac34$.

\subsection{Interacting SAWs}

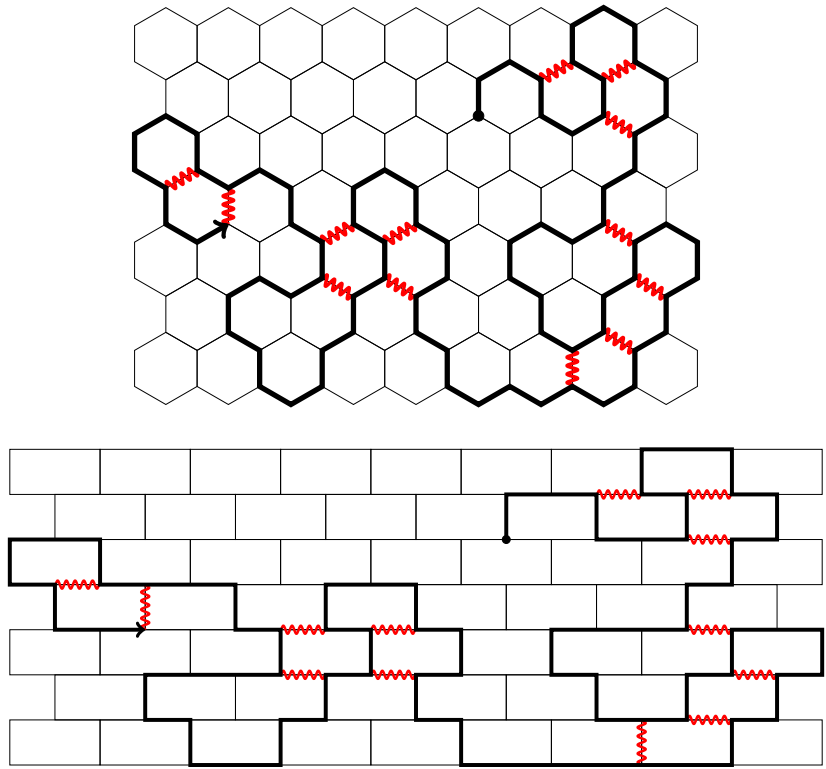
\begin{figure}[ht]
\begin{subfigure}{\textwidth}
\centering
\scalebox{0.8}{
\begin{tikzpicture}[rotate=180]
\def\xs{0.519615} \def\ys{0.3}

\foreach \x in {0,...,8}
\foreach \y in {0,...,3}
{
\draw[xshift=2*\xs*\x cm,yshift=6*\ys*\y cm] (0,0)--(0,2*\ys)--(\xs,3*\ys)--(2*\xs,2*\ys)--(2*\xs,0)--(\xs,-\ys)--cycle;
}

\foreach \x in {0,...,7}
\foreach \y in {0,...,2}
{
\draw[xshift=\xs cm +2*\xs*\x cm,yshift= 3*\ys cm+6*\ys*\y cm] (0,0)--(0,2*\ys)--(\xs,3*\ys)--(2*\xs,2*\ys)--(2*\xs,0)--(\xs,-\ys)--cycle;
}

\draw[ultra thick,red,decorate,decoration=snake,segment length=4] (5*\xs,3*\ys) -- (4*\xs,2*\ys);
\draw[ultra thick,red,decorate,decoration=snake,segment length=4] (3*\xs,3*\ys) -- (2*\xs,2*\ys);
\draw[ultra thick,red,decorate,decoration=snake,segment length=4] (3*\xs,5*\ys) -- (2*\xs,6*\ys);
\draw[ultra thick,red,decorate,decoration=snake,segment length=4] (3*\xs,11*\ys) -- (2*\xs,12*\ys);
\draw[ultra thick,red,decorate,decoration=snake,segment length=4] (2*\xs,14*\ys) -- (\xs,15*\ys);
\draw[ultra thick,red,decorate,decoration=snake,segment length=4] (3*\xs,17*\ys) -- (2*\xs,18*\ys);
\draw[ultra thick,red,decorate,decoration=snake,segment length=4] (4*\xs,18*\ys) -- (4*\xs,20*\ys);
\draw[ultra thick,red,decorate,decoration=snake,segment length=4] (10*\xs,14*\ys) -- (9*\xs,15*\ys);
\draw[ultra thick,red,decorate,decoration=snake,segment length=4] (10*\xs,12*\ys) -- (9*\xs,11*\ys);
\draw[ultra thick,red,decorate,decoration=snake,segment length=4] (12*\xs,14*\ys) -- (11*\xs,15*\ys);
\draw[ultra thick,red,decorate,decoration=snake,segment length=4] (12*\xs,12*\ys) -- (11*\xs,11*\ys);
\draw[ultra thick,red,decorate,decoration=snake,segment length=4] (16*\xs,8*\ys) -- (17*\xs,9*\ys);
\draw[ultra thick,red,decorate,decoration=snake,segment length=4] (15*\xs,11*\ys) -- (15*\xs,9*\ys);

\filldraw (7*\xs,5*\ys) circle (2.5pt);

\draw[line width=2.5pt,->]  (7*\xs,5*\ys)-- ++(0,-2*\ys)-- ++(-\xs,-\ys)-- ++(-\xs,\ys)-- ++(0,2*\ys)-- ++(-\xs,\ys)-- ++(-\xs,-\ys)-- ++(0,-2*\ys)-- ++(\xs,-\ys)-- ++(0,-2*\ys)
-- ++(-\xs,-\ys)-- ++(-\xs,\ys)-- ++(0,2*\ys)-- ++(-\xs,\ys)-- ++(0,2*\ys)-- ++(\xs,\ys)-- ++(0,2*\ys)-- ++(\xs,\ys)-- ++(0,2*\ys)-- ++(\xs,\ys)-- ++(\xs,-\ys)-- ++(\xs,\ys)-- ++(0,2*\ys)
-- ++(-\xs,\ys)-- ++(0,2*\ys)-- ++(-\xs,\ys)-- ++(-\xs,-\ys)-- ++(0,-2*\ys)-- ++(-\xs,-\ys)-- ++(0,-2*\ys)-- ++(-\xs,-\ys)-- ++(-\xs,\ys)-- ++(0,2*\ys)-- ++(\xs,\ys)-- ++(0,2*\ys)
-- ++(\xs,\ys)-- ++(0,2*\ys)-- ++(\xs,\ys)-- ++(\xs,-\ys)-- ++(\xs,\ys)-- ++(\xs,-\ys)-- ++(\xs,\ys)-- ++(\xs,-\ys)-- ++(0,-2*\ys)-- ++(\xs,-\ys)-- ++(0,-2*\ys)-- ++(-\xs,-\ys)
-- ++(0,-2*\ys)-- ++(\xs,-\ys)-- ++(0,-2*\ys)-- ++(\xs,-\ys)-- ++(\xs,\ys)-- ++(0,2*\ys)-- ++(-\xs,\ys)-- ++(0,2*\ys)-- ++(\xs,\ys)-- ++(0,2*\ys)-- ++(\xs,\ys)-- ++(0,2*\ys)
-- ++(\xs,\ys)-- ++(\xs,-\ys)-- ++(0,-2*\ys)-- ++(\xs,-\ys)-- ++(0,-2*\ys)-- ++(-\xs,-\ys)-- ++(-\xs,\ys)-- ++(-\xs,-\ys)-- ++(0,-2*\ys)-- ++(\xs,-\ys) -- ++(0,-2*\ys)
-- ++(\xs,-\ys)-- ++(\xs,\ys)-- ++(\xs,-\ys)-- ++(0,-2*\ys) -- ++(\xs,-\ys)-- ++(\xs,\ys)-- ++(0,2*\ys)-- ++(-\xs,\ys)-- ++(0,2*\ys)-- ++(-\xs,\ys) -- ++(-\xs,-\ys);
\end{tikzpicture}
}
\end{subfigure}

\vspace{0.5cm}

\begin{subfigure}{\textwidth}
\centering
    \scalebox{0.6}{
    \begin{tikzpicture}
    
    \foreach \x in {0,...,8}
    \foreach \y in {0,...,3}
    {
    \draw[xshift=2*\x cm,yshift=2*\y cm] (0,0)--(0,1)--(2,1)--(2,0)--cycle;
    }

    \foreach \x in {0,...,7}
    \foreach \y in {0,...,2}
    {
    \draw[xshift=1 cm +2*\x cm,yshift= 1 cm+2*\y cm] (0,0)--(0,1)--(2,1)--(2,0)--cycle;
    }
    
    \draw[ultra thick, red, decorate, decoration=snake, segment length=5.5] (1,4) -- (2,4);
    \draw[ultra thick, red, decorate, decoration=snake, segment length=5.5] (3,3) -- (3,4);
    \draw[ultra thick, red, decorate, decoration=snake, segment length=5.5] (6,3) -- (7,3);
    \draw[ultra thick, red, decorate, decoration=snake, segment length=5.5] (6,2) -- (7,2);
    \draw[ultra thick, red, decorate, decoration=snake, segment length=5.5] (8,3) -- (9,3);
    \draw[ultra thick, red, decorate, decoration=snake, segment length=5.5] (8,2) -- (9,2);
    \draw[ultra thick, red, decorate, decoration=snake, segment length=5.5] (14,0) -- (14,1);
    \draw[ultra thick, red, decorate, decoration=snake, segment length=5.5] (15,1) -- (16,1);
    \draw[ultra thick, red, decorate, decoration=snake, segment length=5.5] (16,2) -- (17,2);
    \draw[ultra thick, red, decorate, decoration=snake, segment length=5.5] (15,3) -- (16,3);
    \draw[ultra thick, red, decorate, decoration=snake, segment length=5.5] (15,5) -- (16,5);
    \draw[ultra thick, red, decorate, decoration=snake, segment length=5.5] (15,6) -- (16,6);
    \draw[ultra thick, red, decorate, decoration=snake, segment length=5.5] (13,6) -- (14,6);

    \filldraw (11,5) circle (2.5pt);
    
    \draw[line width=2.5pt,->] (11,5) -- ++(0,1) -- ++(1,0) -- ++(1,0) -- ++(0,-1) -- ++(1,0) -- ++(1,0) -- ++(0,1) -- ++(-1,0) -- ++(0,1) -- ++(1,0) -- ++(1,0) -- ++(0,-1) -- ++(1,0) -- ++(0,-1) -- ++(-1,0) -- ++(0,-1) -- ++(-1,0) -- ++(0,-1) -- ++(-1,0) -- ++(-1,0) -- ++(-1,0) -- ++(0,-1) -- ++(1,0) -- ++(0,-1) -- ++(1,0) -- ++(1,0) -- ++(0,1) -- ++(1,0) -- ++(0,1) -- ++(1,0) -- ++(1,0) -- ++(0,-1) -- ++(-1,0) -- ++(0,-1) -- ++(-1,0) -- ++(0,-1) -- ++(-1,0) -- ++(-1,0) -- ++(-1,0) -- ++(-1,0) -- ++(-1,0) -- ++(-1,0) -- ++(0,1) -- ++(-1,0) -- ++(0,1) -- ++(1,0) -- ++(0,1) -- ++(-1,0) -- ++(0,1) -- ++(-1,0) -- ++(-1,0) -- ++(0,-1) -- ++(1,0) -- ++(0,-1) -- ++(-1,0) -- ++(0,-1) -- ++(-1,0) -- ++(0,-1) -- ++(-1,0) -- ++(-1,0) -- ++(0,1) -- ++(-1,0) -- ++(0,1) -- ++(1,0) -- ++(1,0) -- ++(1,0) -- ++(0,1) -- ++(-1,0) -- ++(0,1) -- ++(-1,0) -- ++(-1,0) -- ++(-1,0) -- ++(0,1) -- ++(-1,0) -- ++(-1,0) -- ++(0,-1) -- ++(1,0) -- ++(0,-1) -- ++(1,0) -- ++(1,0);

    \end{tikzpicture}
    }
    \end{subfigure}
\caption{Top: An example of an ISAW on a $7\times 9$ rectangular section of the honeycomb lattice. 
This ISAW has length 80 and contains 13 contacts. Bottom: The same ISAW, embedded in our chosen brickwork lattice.
} 
\label{fig:hcisaw}
\end{figure}

SAWs serve as a model for dilute polymers in a good solvent, but in order to model compact or collapsing polymers we need to introduce an energy term. We let energy $\epsilon$ be associated with two vertices $\omega_i$ and $\omega_j$ if $|\omega_i-\omega_j|=1$ and $|i-j|>1$; when this occurs the walk is said to have a \emph{nearest neighbour contact} or just \emph{contact}. If $c(\omega)$ is the total number of contacts in $\omega$ then we let $\cw=\exp(-\epsilon/kT)$ where $k$ is Boltzmann's constant and $T$ is absolute temperature, and assign weight $\cw^{c(\omega)}$ to $\omega$. SAWs which have been so weighted are \emph{interacting} SAWs or ISAWs. An example of an ISAW is shown in Figure~\ref{fig:hcisaw}.

The \emph{partition function} for ISAWs of length $n$ is 
\begin{equation}
    c_n(\cw) = \sum_{\omega\in\bO_n}\cw^{c(\omega)} = \sum_m c_{n,m}\cw^m
\end{equation}
where $c_{n,m}$ is the number of ISAWs of length $n$ with $m$ contacts. Note that it is expected that the limiting free energy
\begin{equation}\label{eqn:defn_kappa}
    \kappa(\cw) = \lim_{n\to\infty} \frac1n \log c_n(\cw)
\end{equation}
exists for all $\cw\geq0$, but this has only been proved for $0\leq\cw\leq1$ (see eg.~\cite[Chapter 9]{BvR2015}). Define the bivariate generating function for honeycomb lattice ISAWs as
\begin{align}
\CGf(\cw,\ew) &= \sum_{m,n} c_{m,n} \cw^m \ew^n \notag \\
&= 1+ 3\ew + 6\ew^2+ 12\ew^3 +24\ew^4 + (42+\cw)\ew^5 + (78+12\cw)\ew^6 +\notag \\
 &\hspace{1cm}  (144+30\cw)\ew^7 + (264+72\cw)\ew^8+(486 +126\cw+36\cw^2)\ew^9 + \cdots
\end{align}

As $\cw$ is varied the behaviour of an `average' ISAW of length $n$ (sampled from the Boltzmann distribution, where $\mathbb{P}_n(\omega) \propto \cw^{c(\omega)}$) changes. The mean metric quantities defined in~\eqref{eqn:mean_metric_quantities} can now be taken as functions of $\cw$:
\begin{equation}
    \ave{R^2_x}_n = \frac{1}{c_n(\cw)}\sum_{\omega\in\bO_n}R^2_x(\omega)\cw^{c(\omega)}.
\end{equation}

For small $\cw$ (corresponding to high temperature), the polymers are in the dilute phase and the exponents $\gamma$ and $\nu$ are expected to be independent of $\cw$ and the same as the $\cw=1$ case. For large $\cw$ (low temperature) the average ISAW is highly compact, corresponding to polymers in a poor solvent. For two dimensions, the exponents $\gamma$ and $\nu$ in the compact phase are expected to be $\frac{19}{16}$ and $\frac12$ respectively~\cite{DS87dense}. The \emph{$\theta$-temperature} (corresponding to a particular value of $\cw$, which we call $\cwc$) separates these two regimes. 

It is generally accepted now that at the $\theta$-point, the exponents $\gamma$ and $\nu$ take values $\gamma=\frac87\approx 1.143$ and $\nu=\frac47\approx 0.571$, respectively. These values were originally obtained by Duplantier and Saleur~\cite{DS87}, who studied a variant of ISAWs on the honeycomb lattice with annealed vacancies, where the weight is effectively associated with nearest-neighbour and (some) next-nearest-neighbour contacts. They argued that this model was in the same universality class as regular ISAWs. Various numerical studies (eg.~\cite{CGPP, Gherardi2013, LKL2011}) have found estimates consistent with these values.

There has been some confusion around these exponents, however. Nienhuis and collaborators~\cite{BN1989,WBN1992} studied an exactly solvable $\bigO(n)$ loop model on the square lattice, where for a certain set of vertex weights the $n\to0$ limit was proposed to be in the same universality class as ISAWs (this is now called the BN model, after Bl\"{o}te and Nienhuis). This led to the conclusion that at the $\theta$-point, $\gamma=\frac{53}{46}\approx 1.152$ and $\nu=\frac{12}{23}\approx 0.522$. Subsequent numerical work on that model~\cite{BOP2013} produced estimates $\gamma\approx1.045$ and $\nu\approx0.576$, much closer to the Duplantier-Saleur values. An explanation was offered by Vernier et al.~\cite{VJS2015}, who argued that the BN model is actually in a completely different universality class from ISAWs, with critical exponents that form a continuum rather than taking on discrete values. The numerical discrepancy is then explained by the fact that the BN values are the lower bounds for the range of exponents.

\subsection{Amplitude ratios}

For non-interacting SAWs, the metric properties described above are related by the CSCPS equality, proposed by Cardy and Saleur \cite{CS89}, queried as incompatible with enumerations by Guttmann and Yang \cite{GY90} and subsequently corrected by Caracciolo, Pelissetto and Sokal \cite{CPS90}. Let 
\begin{equation}
    A_n=\frac{\ave{\Rgs}_n}{\ave{\Res}_n},\qquad B_n=\frac{\ave{\Rms}_n}{\ave{\Res}_n}.    
\end{equation}
The equality is
\begin{equation}
    \lim_{n \to \infty} F_n = 0,
\end{equation}
where
\begin{equation} \label{FN}
    F_n=\left ( 2+\frac{y_t}{y_h}\right) A_n-2B_n+\frac{1}{2},
\end{equation}
and $y_t = \frac{4}{3}$ and  $y_h = \frac{91}{48}$ are the thermal and magnetic renormalisation-group eigenvalues, respectively, of the O$(0)$ model, that is to say, the SAW model. For brevity we will sometimes write $A_\infty = \lim_{n\to\infty} A_n$ and $B_\infty = \lim_{n\to\infty} B_n$

For the interacting SAW model at the $\theta$-point, Owczarek et al.~\cite{OPBG} conjectured that the CSCPS equality holds with $y_t$ and $y_h$ in (\ref{FN}) replaced by their $\theta$-point values, notably $y_t = \frac{7}{4}$ and  $y_h = 2.$ Of course, the metric properties must also be calculated at the $\theta$-point. In \cite{OPBG} compelling numerical evidence for this conjecture was presented, but no proof. In the intervening 25 years there has still been no proof of this result, but in 2011 Caracciolo et al.~\cite{CGPP} presented an exhaustive Monte Carlo analysis which provided even more compelling numerical evidence for the correctness of this conjecture.

\subsection{Honeycomb ISAWs and generating function identities}

It is very tempting to try to use the methods of Duminil-Copin and Smirnov~\cite{DS12} and Beaton et al.~\cite{BBDDG} to compute and prove the exact value of $\cwc$. There are, however, (at least) two issues with this idea, which we briefly describe here.

In their proof of the connective constant, Duminil-Copin and Smirnov show that an identity involving weighted SAW generating functions is satisfied for particular values of the weights. More precisely, they take a finite, simply connected region $\mathcal{D}$ of the lattice, and for convenience let SAWs start and end on the \emph{midpoints} of edges. Then for $a$ a midpoint on the boundary $\partial\mathcal{D}$ and $b$ any midpoint, they define the \emph{parafermionic observable}
\begin{equation}\label{eqn:parafermionic_observable}
    F_{a\to b}(\ew,\sigma) = \sum_{\omega:a\to b}e^{-i\sigma W(\omega)}\ew^{|\omega|},
\end{equation}
where the sum is over all SAWs in $\mathcal{D}$ from $a$ to $b$, $|\omega|$ is the length of $\omega$, and $W(\omega)$ is the winding angle of $\omega$ (that is, $\frac{\pi}{3}$ times the number of left turns minus the number of right turns).

The identity is then
\begin{equation}\label{eqn:pf_local_identity}
\textstyle
    (p-v)F_{a\to p}(\ewc,\frac58) + (q-v)F_{a\to q}(\ewc,\frac58) + (r-v)F_{a\to r}(\ewc,\frac58) = 0,
\end{equation}
where $p,q,r$ are the three midpoints around a vertex $v$, $a$ is any midpoint in $\partial\mathcal{D}$, and $\ewc=1/\sqrt{2+\sqrt{2}}$. This identity is proved by partitioning walks which end adjacent to $v$ into two sets: those which visit at most two of $p,q,r$, and those which visit all three. In the first (resp.~second) set, the contributions of triples (resp.~pairs) of walks to~\eqref{eqn:pf_local_identity} can then be shown to be 0.

To accommodate ISAWs, one would need to generalise~\eqref{eqn:parafermionic_observable} by including the weight $u$ for nearest-neighbour contacts, and hope that~\eqref{eqn:pf_local_identity} (or something like it) could be satisfied for values of $\cw$ other than the trivial $\cw=1$ case. Unfortunately we have not been able to construct such an identity, and would be surprised if such a thing exists. This is the first problem.

Even if an identity like~\eqref{eqn:pf_local_identity} could be found, there is another issue. The proof in~\cite{BBDDG} for the critical surface adsorption fugacity $a_\mathrm{c}$ relies on taking the generating function $\mathcal{G}_T(a,\ew)$ for SAWs in an infinite strip of width $T$, with $a$ tracking the number of visits to one of the sides of the strip. One can then set $\ew=\ewc$ and interpret this as a series in $a$, rather than $\ew$. This series has a radius of convergence, say $a_T$, and it is shown in~\cite{BBDDG} that $a_T$ decreases to $a_\mathrm{c}$ as $T\to\infty$. This works because $a_\mathrm{c}$ is the largest value of $a$ such that the radius of convergence of $\mathcal{G}(a,\ew)$ (once more viewed as a series in $\ew$) is $\ewc$, where $\mathcal{G}(a,\ew)$ is the generating function for TAWs which accrue weight $a$ with each visit to the boundary.

\begin{figure}[ht]
\centering
\begin{tikzpicture}
\node at (0,0) [anchor=south west] {\includegraphics[width=0.5\textwidth]{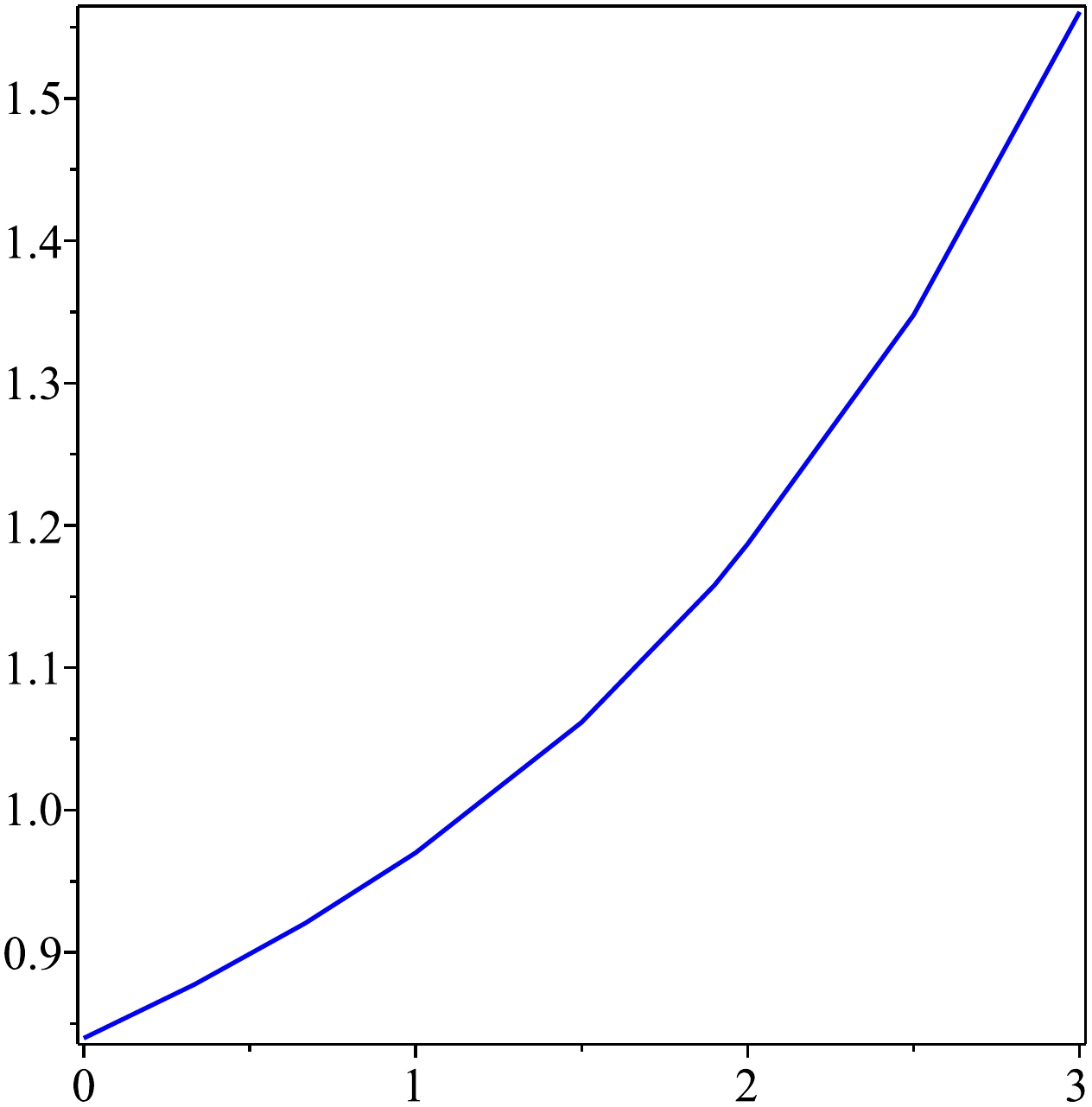}};
\node at (-0.4,4.5) {$\kappa(u)$};
\node at (4.5,-0.2) {$u$};
\end{tikzpicture}
 \caption{Square lattice. Plot of $\kappa(u)$ against $u$.}
 \label{fig:sqfe}
\end{figure}

However, this is not true for the ISAW generating function $\mathcal{C}(\cw,\ew)$. Because $\kappa(u)$ (as defined in~\eqref{eqn:defn_kappa}) is expected to be monotone increasing in $\cw$, the only value of $\cw$ for which $\ewc$ is the radius of convergence of $\mathcal{C}(\cw,\ew)$ is $\cw=1$. Instead, at $\cw=\cwc$, the radius of convergence will be some other value smaller than $\ewc$, about which we presently have no further exact information. In Figure~\ref{fig:sqfe} we show a plot of $\kappa(u)$ against $u,$ derived from the series data we discuss below.

\subsection{ISAWs in three dimensions}
One interesting aspect of ISAWs is that $d=3$ is the critical dimension.
In \cite{BLS19} two related models on ${\mathbb Z}^3$ are studied. One is a model of classical unbounded $n$-component continuous spins with a triple-well single-spin potential (the $|\phi|^6$ model), and the other is a random walk model of linear polymers with a three-body repulsion and two-body attraction at the tricritical theta point. The polymer model is exactly equivalent to a supersymmetric spin model which corresponds to the $n=0$ version of the $|\phi|^6$ model. For both models the authors of~\cite{BLS19} identify the tricritical point, and prove that the tricritical two-point function has Gaussian long-distance decay, namely $1/|x|.$

Recently Bauerschmidt and Slade \cite{BS19} studied the full phase diagram for the mean field version of this problem, by considering walks on the complete graph in the limit as the number of vertices tends to infinity.
In this very accessible account, they find a dilute phase which is separated from a dense phase by a phase boundary curve. The phase boundary is divided into two parts, corresponding to first-order and second-order phase transitions, with the division occurring at a tricritical point.

\subsection{Outline of the paper} The primary goal of this paper is to analyse ISAW series data in order to compute new estimates for the amplitude ratios, critical points and critical exponents described above. In Section~\ref{sec:series} we outline the finite lattice method and its application to enumerating ISAWs. The numerical analysis of the series data is given in Section~\ref{sec:results}. Section~\ref{ssec:non_interact} reviews some known results for non-interacting SAWs. In Section~\ref{sssec:interact_hc} we compute several estimates for $\cwc$ for the honeycomb lattice, and in Section~\ref{sssec:interact_sq} this is repeated for the square lattice. Section~\ref{ssec:bridges_taws} presents some analysis of the critical behaviour of interacting bridges and TAWs on the square lattice. Finally in Section~\ref{sec:conclusion} we present some concluding remarks.

\section{Series generation}\label{sec:series}

\subsection{Hexagonal lattice}
We calculated series for the ISAW generating function $\CGf(\cw,\ew)$ for walks of up to 75 steps.

\subsubsection{Enumeration of ISAW \label{sec:flm}}

The algorithm we use to enumerate ISAWs on the honeycomb lattice builds on the 
pioneering work of Enting \cite{IGE80e} who enumerated square lattice 
self-avoiding polygons using the finite lattice method. An implementation of the SAP enumeration 
algorithm for the honeycomb lattice can be found in \cite{EG89a}.  The basic idea of the finite 
lattice method is to calculate partial generating functions for various properties 
of a given model on finite pieces, say $W \times L$ rectangles of the  given 
lattice, and then reconstruct a series expansion for the infinite lattice
limit by combining the results from the finite pieces. The generating
function for any finite piece is calculated using transfer matrix (TM)
techniques. Our algorithm is based in large part on the one devised by Conway, Enting and 
Guttmann \cite{CEG93} for the enumeration of square lattice SAWs with various
improvements by Jensen \cite{IJ04}. The TM algorithm for the enumeration
of honeycomb SAW was described in \cite{IJ06} and a detailed description
of the general method can be found in \cite{PolygonBook}.

\subsubsection{Basic transfer matrix algorithm}

\begin{figure}[ht]
\begin{center}

\begin{tikzpicture}[scale=0.8]  
\def\xs{0.519615} \def\ys{0.3}

\foreach \x in {0,...,6}
\foreach \y in {0,...,2}
{
\draw[xshift=2*\xs*\x cm,yshift=6*\ys*\y cm] (0,0)--(0,2*\ys)--(\xs,3*\ys)--(2*\xs,2*\ys)--(2*\xs,0)--(\xs,-\ys)--cycle;
}

\foreach \x in {0,...,5}
\foreach \y in {0,...,1}
{
\draw[xshift=\xs cm +2*\xs*\x cm,yshift= 3*\ys cm+6*\ys*\y cm] (0,0)--(0,2*\ys)--(\xs,3*\ys)--(2*\xs,2*\ys)--(2*\xs,0)--(\xs,-\ys)--cycle;
}
\draw[very thick] (4.5*\xs,-2*\ys)--(4.5*\xs,7*\ys)--(5.5*\xs,7*\ys)--(5.5*\xs,16*\ys);
\draw[very thick,dotted]  (4.5*\xs,\ys)--(5.5*\xs,\ys)--(5.5*\xs,7*\ys);

\draw[line width=2.5pt]  (6*\xs,12*\ys)--++(-\xs,-\ys)--++(0,-2*\ys)--++(-\xs,-\ys)--++(-\xs,\ys);
\draw[line width=2.5pt]  (5*\xs,5*\ys)--++(-\xs,\ys)--++(-\xs,-\ys)--++(-\xs,\ys)--++(-\xs,-\ys)--++(0,-2*\ys)--++(\xs,-\ys)--++(\xs,\ys)--++(\xs,-\ys)--++(0,-2*\ys)--++(\xs,-\ys);
\draw[ultra thick,red,decorate,decoration=snake,segment length=4] (4*\xs,6*\ys) -- (4*\xs,8*\ys);
\draw[ultra thick,red,decorate,decoration=snake,segment length=4] (3*\xs,3*\ys) -- (3*\xs,5*\ys);

\begin{scope}[xshift=9cm,yshift=-1.0cm,rotate=0]
\foreach \x in {0,...,4}
\foreach \y in {0,...,3}
{
\draw[xshift=2*\xs*\x cm,yshift=6*\ys*\y cm]  (0,0)--(0,2*\ys)--(\xs,3*\ys)--(2*\xs,2*\ys)--(2*\xs,0)--(\xs,-\ys)--cycle;
}

\foreach \x in {0,...,3}
\foreach \y in {0,...,2}
{
\draw[xshift=\xs cm +2*\xs*\x cm,yshift= 3*\ys cm+6*\ys*\y cm]  (0,0)--(0,2*\ys)--(\xs,3*\ys)--(2*\xs,2*\ys)--(2*\xs,0)--(\xs,-\ys)--cycle;
}
\draw[very thick]  (-\xs,13*\ys)--(3.5*\xs,13*\ys)--(3.5*\xs,10*\ys)--(11*\xs,10*\ys);
\draw[very thick,dotted]  (3.5*\xs,13*\ys)--(5.5*\xs,13*\ys)--(5.5*\xs,10*\ys);
 
\end{scope}

\end{tikzpicture}  

\end{center}
\caption{\label{fig:transfer}
A snapshot of the boundary line (solid line) during the transfer matrix 
calculation on the honeycomb lattice. ISAWs are enumerated by successive
moves of the kink in the boundary line, as exemplified by the position given 
by the dotted line, so that two vertices at a time are added to the rectangle. 
The left panel show the case $L\geq W$ while the right panel is the case $L<W$.
To the left of the boundary line in the left panel we have drawn an example of a 
partially completed ISAW.}
\end{figure}
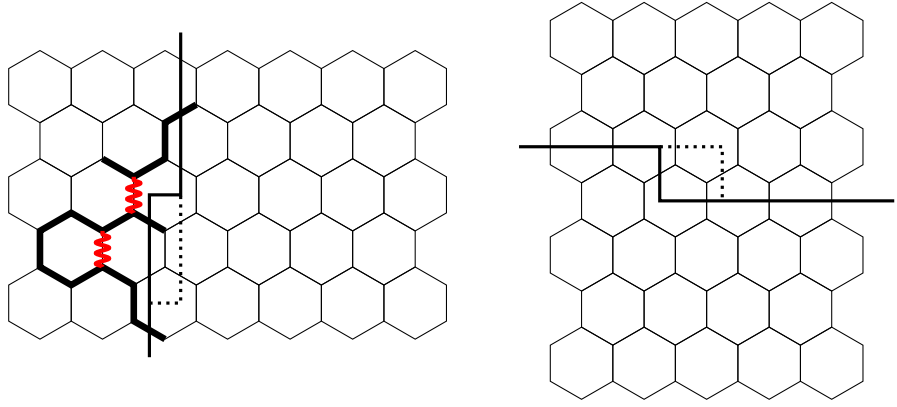

The most efficient implementation of the TM algorithm generally involves 
bisecting the finite lattice with a boundary line and moving this 
boundary in such a way as to build up the lattice cell by cell. 
 ISAW in a given rectangle are enumerated by moving the 
boundary so as to add  two vertices at a time, as illustrated in 
Figure~\ref{fig:transfer}. Due to the symmetries of the honeycomb lattice we 
 separately consider  rectangles with  $L < W$ and $L \geq W$, since they 
 will be traversed differently. The reason we do this is that we need to
 minimise the number of edges intersected by the boundary.
 
For each configuration of occupied or empty edges 
along the boundary we maintain a generating function for partial ISAW 
cutting the intersection in that particular pattern. If we draw an ISAW and 
then cut it by a line, we observe that the partial ISAW to the left of the 
boundary line consists of loops connecting two edges on the boundary (we shall refer to 
these as loop-ends),  and pieces  connected to 
only one edge  on the boundary (we call these free ends). The other end of a free piece is 
an end-point of the ISAW so there are at most two free ends.  Furthermore the
number of free ends increase as the boundary is moved.

We are not allowed to form closed loops, so two loop ends can only be joined 
if they belong to different loops. To exclude loops which close on themselves 
we label the occupied edges in such a way that we can easily determine 
whether or not two loop ends belong to the same loop. On two-dimensional 
lattices this can be done by relying on the fact that two loops 
can never intertwine. Each loop-edge is assigned a label
depending on whether it is the lower or upper end of a loop. We must also ensure
that the graphs we are counting have only a single component. In addition
we demand that valid ISAW configurations are those where the walk touches
all the sides of the rectangle.

Unoccupied edges come in two varieties depending on whether or not
the ISAW passes through the vertex to the left of the edge. If
this `left' vertex is empty no contact can occur along this edge but
if the vertex is occupied a contact may occur depending on whether
or not the ISAW passes through the rightmost vertex along the edge in question. 
This is illustrated in Figure~\ref{fig:transfer} where for example the second edge along the
boundary is a `contact' edge since the ISAW passes through the vertex
on the left while the top-most edge in an `empty' edge since the vertex
to the left is unoccupied.

 Each
configuration along the boundary line can thus be represented by a set of 
edge states $\{\sigma_i\}$, where

\begin{equation}\label{eq:states}
\sigma_i  = \left\{ \begin{array}{rl}
\ee &\;\;\; \mbox{empty edge},  \\ 
\ec &\;\;\; \mbox{contact edge},  \\ 
\el &\;\;\; \mbox{lower loop end}, \\
\eu &\;\;\; \mbox{upper loop end}, \\
\ef &\;\;\; \mbox{free end}. \\
\end{array} \right.
\end{equation}
\noindent
If we read from the bottom to the top, the configuration along the 
intersection of the partial ISAW in Figure~\ref{fig:transfer} is $\{\el\ec\eu\ec\ef\ee\}$. 
 
\subsubsection{Updating rules for the TM algorithm}
 
The updating of a partial generating function depends on the states of the
edges to the left of the new vertices. When the kink in the boundary is moved
we insert 3 new edges namely the edge connecting the new vertices and the
two edges to their right.  In Fig~\ref{fig:update} we display the possible local 
`input' states and the `output' states which arise as the kink in the boundary
is propagated by one step. Not all 25 possible local input states are displayed
since many are related by an obvious reflection symmetry while others
are related by the interchange $\el \leftrightarrow \eu$ with straightforward changes
to the corresponding updating rules. 
 We shall refer to the configuration before the move 
as the `source' and a configuration produced as a result of the move as a `target'.
In each move the source generating function is multiplied by $\cw^j\ew^k$,
where $j$ is the number of added contacts and $k$ is the number of new occupied edges 
and is then added to the target generating function.

\begin{figure}[ht]
\begin{center}
\scalebox{0.85}{
\begin{tikzpicture}

\scriptsize

\def\xs{0.519615} \def\ys{0.3}

\begin{scope}[yshift=24 cm]
\foreach \x in {0,...,6} {
\draw[xshift= 2.5*\x cm] (-\xs,-\ys)--(0,0)--(\xs,-\ys) (-\xs,3*\ys)--(0,2*\ys)--(\xs,3*\ys) (0,0)--(0,2*\ys);
\node[xshift= 2.5*\x cm] at (-1.3*\xs,-\ys) {\ee}; \node[xshift= 2.5*\x cm] at (-1.3*\xs,3*\ys) {\ee}; 
}

\node  at (1.3*\xs,-\ys) {\ee}; \node at (1.3*\xs,3*\ys) {\ee}; \node at (0,-2*\ys) {\normalsize $1$}; 

\begin{scope}[xshift=2.5 cm]
\draw[ultra thick]  (\xs,-\ys)--(0,0)--(0,2*\ys)--(\xs,3*\ys);
\node  at (1.3*\xs,-\ys) {\el}; \node at (1.3*\xs,3*\ys) {\eu}; \node at (0,-2*\ys) {\normalsize $\ew^3$};
\end{scope}

\begin{scope}[xshift=5 cm]
\draw[ultra thick]  (\xs,-\ys)--(0,0) (0,2*\ys)--(\xs,3*\ys);
\draw[very thick,red,decorate,decoration=snake,segment length=4] (0,0) -- (0,2*\ys);
\filldraw (0,0) circle (2pt) (0,2*\ys) circle (2pt);
\node at (1.3*\xs,-\ys) {\ef}; \node at (1.3*\xs,3*\ys) {\ef}; \node at (0,-2*\ys) {\normalsize $\cw\ew^2$};
\end{scope}

\begin{scope}[xshift=7.5 cm]
\draw[ultra thick]  (\xs,-\ys)--(0,0) ;
\filldraw (0,0) circle (2pt);
\node  at (1.3*\xs,-\ys) {\ef}; \node at (1.3*\xs,3*\ys) {\ee}; \node at (0,-2*\ys) {\normalsize $\ew$};
\end{scope}

\begin{scope}[xshift=10 cm]
\draw[ultra thick]  (\xs,-\ys)--(0,0)--(0,2*\ys) ;
\filldraw  (0,2*\ys) circle (2pt);
\node at (1.3*\xs,-\ys) {\ef}; \node at (1.3*\xs,3*\ys) {\ec}; \node at (0,-2*\ys) {\normalsize $\ew^2$};
\end{scope}

\begin{scope}[xshift=12.5 cm]
\draw[ultra thick]  (\xs,3*\ys)--(0,2*\ys) ;
\filldraw  (0,2*\ys) circle (2pt);
\node at (1.3*\xs,-\ys) {\ee}; \node at (1.3*\xs,3*\ys) {\ef};  \node at (0,-2*\ys) {\normalsize $\ew$};
\end{scope}

\begin{scope}[xshift=15 cm]
\draw[ultra thick]  (\xs,3*\ys)--(0,2*\ys)--(0,0) ;
\filldraw (0,0) circle (2pt);
\node at (1.3*\xs,-\ys) {\ec}; \node at (1.3*\xs,3*\ys) {\ef};  \node at (0,-2*\ys) {\normalsize $\ew^2$};
\end{scope}
\end{scope}

\begin{scope}[yshift=21 cm]
\foreach \x in {0,...,6} {
\draw[xshift= 2.5*\x cm] (-\xs,-\ys)--(0,0)--(\xs,-\ys) (-\xs,3*\ys)--(0,2*\ys)--(\xs,3*\ys) (0,0)--(0,2*\ys);
\node[xshift= 2.5*\x cm] at (-1.3*\xs,-\ys) {\ee}; \node[xshift= 2.5*\x cm] at (-1.3*\xs,3*\ys) {\ec}; 
}

\node  at (1.3*\xs,-\ys) {\ee}; \node at (1.3*\xs,3*\ys) {\ee}; \node at (0,-2*\ys) {\normalsize $1$}; 

\begin{scope}[xshift=2.5 cm]
\draw[ultra thick]  (\xs,-\ys)--(0,0)--(0,2*\ys)--(\xs,3*\ys);
\draw[very thick,red,decorate,decoration=snake,segment length=4] (-\xs,3*\ys) -- (0,2*\ys);
\node  at (1.3*\xs,-\ys) {\el}; \node at (1.3*\xs,3*\ys) {\eu}; \node at (0,-2*\ys) {\normalsize $\cw\ew^3$};
\end{scope}

\begin{scope}[xshift=5 cm]
\draw[ultra thick]  (\xs,-\ys)--(0,0) (0,2*\ys)--(\xs,3*\ys);
\draw[very thick,red,decorate,decoration=snake,segment length=4] (-\xs,3*\ys) -- (0,2*\ys);
\draw[very thick,red,decorate,decoration=snake,segment length=4] (0,0) -- (0,2*\ys);
\filldraw (0,0) circle (2pt) (0,2*\ys) circle (2pt);
\node at (1.3*\xs,-\ys) {\ef}; \node at (1.3*\xs,3*\ys) {\ef}; \node at (0,-2*\ys) {\normalsize $\cw^2\ew^2$};
\end{scope}

\begin{scope}[xshift=7.5 cm]
\draw[ultra thick]  (\xs,-\ys)--(0,0) ;
\filldraw (0,0) circle (2pt);
\node  at (1.3*\xs,-\ys) {\ef}; \node at (1.3*\xs,3*\ys) {\ee}; \node at (0,-2*\ys) {\normalsize $\ew$};
\end{scope}

\begin{scope}[xshift=10 cm]
\draw[ultra thick]  (\xs,-\ys)--(0,0)--(0,2*\ys) ;
\draw[very thick,red,decorate,decoration=snake,segment length=4] (-\xs,3*\ys) -- (0,2*\ys);
\filldraw  (0,2*\ys) circle (2pt);
\node at (1.3*\xs,-\ys) {\ef}; \node at (1.3*\xs,3*\ys) {\ec}; \node at (0,-2*\ys) {\normalsize $\cw\ew^2$};
\end{scope}

\begin{scope}[xshift=12.5 cm]
\draw[ultra thick]  (\xs,3*\ys)--(0,2*\ys) ;
\draw[very thick,red,decorate,decoration=snake,segment length=4] (-\xs,3*\ys) -- (0,2*\ys);
\filldraw   (0,2*\ys) circle (2pt);
\node at (1.3*\xs,-\ys) {\ee}; \node at (1.3*\xs,3*\ys) {\ef};  \node at (0,-2*\ys) {\normalsize $\cw\ew$};
\end{scope}

\begin{scope}[xshift=15 cm]
\draw[ultra thick]  (\xs,3*\ys)--(0,2*\ys)--(0,0) ;
\draw[very thick,red,decorate,decoration=snake,segment length=4] (-\xs,3*\ys) -- (0,2*\ys);
\filldraw (0,0) circle (2pt);
\node at (1.3*\xs,-\ys) {\ec}; \node at (1.3*\xs,3*\ys) {\ef};  \node at (0,-2*\ys) {\normalsize $\cw\ew^2$};
\end{scope}
\end{scope}

\begin{scope}[yshift=18 cm]
\foreach \x in {0,...,6} {
\draw[xshift= 2.5*\x cm] (-\xs,-\ys)--(0,0)--(\xs,-\ys) (-\xs,3*\ys)--(0,2*\ys)--(\xs,3*\ys) (0,0)--(0,2*\ys);
\node[xshift= 2.5*\x cm] at (-1.3*\xs,-\ys) {\ec}; \node[xshift= 2.5*\x cm] at (-1.3*\xs,3*\ys) {\ec}; 
}

\node  at (1.3*\xs,-\ys) {\ee}; \node at (1.3*\xs,3*\ys) {\ee}; \node at (0,-2*\ys) {\normalsize $1$}; 

\begin{scope}[xshift=2.5 cm]
\draw[ultra thick]  (\xs,-\ys)--(0,0)--(0,2*\ys)--(\xs,3*\ys);
\draw[very thick,red,decorate,decoration=snake,segment length=4] (-\xs,3*\ys) -- (0,2*\ys);
\draw[very thick,red,decorate,decoration=snake,segment length=4] (-\xs,-\ys) -- (0,0);
\node  at (1.3*\xs,-\ys) {\el}; \node at (1.3*\xs,3*\ys) {\eu}; \node at (0,-2*\ys) {\normalsize $\cw^2\ew^3$};
\end{scope}

\begin{scope}[xshift=5 cm]
\draw[ultra thick]  (\xs,-\ys)--(0,0) (0,2*\ys)--(\xs,3*\ys);
\draw[very thick,red,decorate,decoration=snake,segment length=4] (-\xs,3*\ys) -- (0,2*\ys);
\draw[very thick,red,decorate,decoration=snake,segment length=4] (0,0) -- (0,2*\ys);
\draw[very thick,red,decorate,decoration=snake,segment length=4] (-\xs,-\ys) -- (0,0);
\filldraw (0,0) circle (2pt) (0,2*\ys) circle (2pt);
\node at (1.3*\xs,-\ys) {\ef}; \node at (1.3*\xs,3*\ys) {\ef}; \node at (0,-2*\ys) {\normalsize $\cw^3\ew^2$};
\end{scope}

\begin{scope}[xshift=7.5 cm]
\draw[ultra thick]  (\xs,-\ys)--(0,0) ;
\draw[very thick,red,decorate,decoration=snake,segment length=4] (-\xs,-\ys) -- (0,0);
\filldraw (0,0) circle (2pt);
\node  at (1.3*\xs,-\ys) {\ef}; \node at (1.3*\xs,3*\ys) {\ee}; \node at (0,-2*\ys) {\normalsize $\cw\ew$};
\end{scope}

\begin{scope}[xshift=10 cm]
\draw[ultra thick]  (\xs,-\ys)--(0,0)--(0,2*\ys) ;
\draw[very thick,red,decorate,decoration=snake,segment length=4] (-\xs,3*\ys) -- (0,2*\ys);
\draw[very thick,red,decorate,decoration=snake,segment length=4] (-\xs,-\ys) -- (0,0);
\filldraw  (0,2*\ys) circle (2pt);
\node at (1.3*\xs,-\ys) {\ef}; \node at (1.3*\xs,3*\ys) {\ec}; \node at (0,-2*\ys) {\normalsize $\cw^2\ew^2$};
\end{scope}

\begin{scope}[xshift=12.5 cm]
\draw[ultra thick]  (\xs,3*\ys)--(0,2*\ys) ;
\draw[very thick,red,decorate,decoration=snake,segment length=4] (-\xs,3*\ys) -- (0,2*\ys);
\filldraw  (0,2*\ys) circle (2pt);
\node at (1.3*\xs,-\ys) {\ee}; \node at (1.3*\xs,3*\ys) {\ef};  \node at (0,-2*\ys) {\normalsize $\cw\ew$};
\end{scope}

\begin{scope}[xshift=15 cm]
\draw[ultra thick]  (\xs,3*\ys)--(0,2*\ys)--(0,0) ;
\draw[very thick,red,decorate,decoration=snake,segment length=4] (-\xs,3*\ys) -- (0,2*\ys);
\draw[very thick,red,decorate,decoration=snake,segment length=4] (-\xs,-\ys) -- (0,0);
\filldraw (0,0) circle (2pt);
\node at (1.3*\xs,-\ys) {\ec}; \node at (1.3*\xs,3*\ys) {\ef};  \node at (0,-2*\ys) {\normalsize $\cw^2\ew^2$};
\end{scope}
\end{scope}

\begin{scope}[yshift=15 cm]
\foreach \x in {0,...,5} {
\draw[xshift= 2.5*\x cm] (-\xs,-\ys)--(0,0)--(\xs,-\ys) (-\xs,3*\ys)--(0,2*\ys)--(\xs,3*\ys) (0,0)--(0,2*\ys);
\draw[xshift= 2.5*\x cm,ultra thick]  (-\xs,-\ys)--(0,0);
\node[xshift= 2.5*\x cm] at (-1.3*\xs,-\ys) {\el}; \node[xshift= 2.5*\x cm] at (-1.3*\xs,3*\ys) {\ee}; 
}

\node  at (1.3*\xs,-\ys) {\ec}; \node at (1.3*\xs,3*\ys) {\ee}; \node at (0,-2*\ys) {\normalsize $1$}; 
\filldraw (0,0) circle (2pt);  \node at (0,4*\ys) {\eu $\, \to$ \ef };

\begin{scope}[xshift=2.5 cm]
\draw[ultra thick]  (0,0)--(0,2*\ys);
\filldraw (0,2*\ys) circle (2pt);
\node at (1.3*\xs,-\ys) {\ec}; \node at (1.3*\xs,3*\ys) {\ec}; \node at (0,-2*\ys) {\normalsize $\ew$};  \node at (0,4*\ys) {\eu $\, \to$ \ef };
\end{scope}

\begin{scope}[xshift=5 cm]
\draw[ultra thick]  (0,0)--(\xs,-\ys);
\node  at (1.3*\xs,-\ys) {\el}; \node at (1.3*\xs,3*\ys) {\ee}; \node at (0,-2*\ys) {\normalsize $\ew$};
\end{scope}

\begin{scope}[xshift=7.5 cm]
\draw[ultra thick]  (0,0)--(0,2*\ys)--(\xs,3*\ys);
\node at (1.3*\xs,-\ys) {\ec}; \node at (1.3*\xs,3*\ys) {\el}; \node at (0,-2*\ys) {\normalsize $\ew^2$};
\end{scope}

\begin{scope}[xshift=10 cm]
\draw[ultra thick]  (0,2*\ys)--(\xs,3*\ys);
\draw[very thick,red,decorate,decoration=snake,segment length=4] (0,0) -- (0,2*\ys);
\filldraw (0,0) circle (2pt) (0,2*\ys) circle (2pt);
\node  at (1.3*\xs,-\ys) {\ec}; \node at (1.3*\xs,3*\ys) {\ef}; \node at (0,-2*\ys) {\normalsize $\cw\ew$};  \node at (0,4*\ys) {\eu $\, \to$ \ef };
\end{scope}

\begin{scope}[xshift=12.5 cm]
\draw[ultra thick] (0,0)--(\xs,-\ys) (\xs,3*\ys)--(0,2*\ys) ;
\draw[very thick,red,decorate,decoration=snake,segment length=4] (0,0) -- (0,2*\ys);
\filldraw  (0,2*\ys) circle (2pt);
\node at (1.3*\xs,-\ys) {\el}; \node at (1.3*\xs,3*\ys) {\ef};  \node at (0,-2*\ys) {\normalsize $\cw\ew^2$}; 
\end{scope}

\end{scope}

\begin{scope}[yshift=12 cm]
\foreach \x in {0,...,5} {
\draw[xshift= 2.5*\x cm] (-\xs,-\ys)--(0,0)--(\xs,-\ys) (-\xs,3*\ys)--(0,2*\ys)--(\xs,3*\ys) (0,0)--(0,2*\ys);
\draw[xshift= 2.5*\x cm,ultra thick]  (-\xs,-\ys)--(0,0);
\node[xshift= 2.5*\x cm] at (-1.3*\xs,-\ys) {\el}; \node[xshift= 2.5*\x cm] at (-1.3*\xs,3*\ys) {\ec}; 
}

\node  at (1.3*\xs,-\ys) {\ec}; \node at (1.3*\xs,3*\ys) {\ee}; \node at (0,-2*\ys) {\normalsize $1$}; 
\filldraw (0,0) circle (2pt);  \node at (0,4*\ys) {\eu $\, \to$ \ef };

\begin{scope}[xshift=2.5 cm]
\draw[ultra thick]  (0,0)--(0,2*\ys);
\draw[very thick,red,decorate,decoration=snake,segment length=4] (-\xs,3*\ys) -- (0,2*\ys);
\filldraw (0,2*\ys) circle (2pt);
\node at (1.3*\xs,-\ys) {\ec}; \node at (1.3*\xs,3*\ys) {\ec}; \node at (0,-2*\ys) {\normalsize $\cw\ew$};  \node at (0,4*\ys) {\eu $\, \to$ \ef };
\end{scope}

\begin{scope}[xshift=5 cm]
\draw[ultra thick]  (0,0)--(\xs,-\ys);
\node  at (1.3*\xs,-\ys) {\el}; \node at (1.3*\xs,3*\ys) {\ee}; \node at (0,-2*\ys) {\normalsize $\ew$};
\end{scope}

\begin{scope}[xshift=7.5 cm]
\draw[ultra thick]  (0,0)--(0,2*\ys)--(\xs,3*\ys);
\draw[very thick,red,decorate,decoration=snake,segment length=4] (-\xs,3*\ys) -- (0,2*\ys);
\node at (1.3*\xs,-\ys) {\ec}; \node at (1.3*\xs,3*\ys) {\el}; \node at (0,-2*\ys) {\normalsize $\cw\ew^2$};
\end{scope}

\begin{scope}[xshift=10 cm]
\draw[ultra thick]  (0,2*\ys)--(\xs,3*\ys);
\draw[very thick,red,decorate,decoration=snake,segment length=4] (-\xs,3*\ys) -- (0,2*\ys);
\draw[very thick,red,decorate,decoration=snake,segment length=4] (0,0) -- (0,2*\ys);
\filldraw (0,0) circle (2pt) (0,2*\ys) circle (2pt);
\node  at (1.3*\xs,-\ys) {\ec}; \node at (1.3*\xs,3*\ys) {\ef}; \node at (0,-2*\ys) {\normalsize $\cw^2\ew$};  \node at (0,4*\ys) {\eu $\, \to$ \ef };
\end{scope}

\begin{scope}[xshift=12.5 cm]
\draw[ultra thick] (0,0)--(\xs,-\ys) (\xs,3*\ys)--(0,2*\ys) ;
\draw[very thick,red,decorate,decoration=snake,segment length=4] (-\xs,3*\ys) -- (0,2*\ys);
\draw[very thick,red,decorate,decoration=snake,segment length=4] (0,0) -- (0,2*\ys);
\filldraw  (0,2*\ys) circle (2pt);
\node at (1.3*\xs,-\ys) {\el}; \node at (1.3*\xs,3*\ys) {\ef};  \node at (0,-2*\ys) {\normalsize $\cw^2\ew^2$}; 
\end{scope}

\end{scope}

\end{tikzpicture}  
}
\end{center}
\caption{\label{fig:update}  Updating rules for the TM enumeration of honeycomb ISAW.
}
\end{figure}
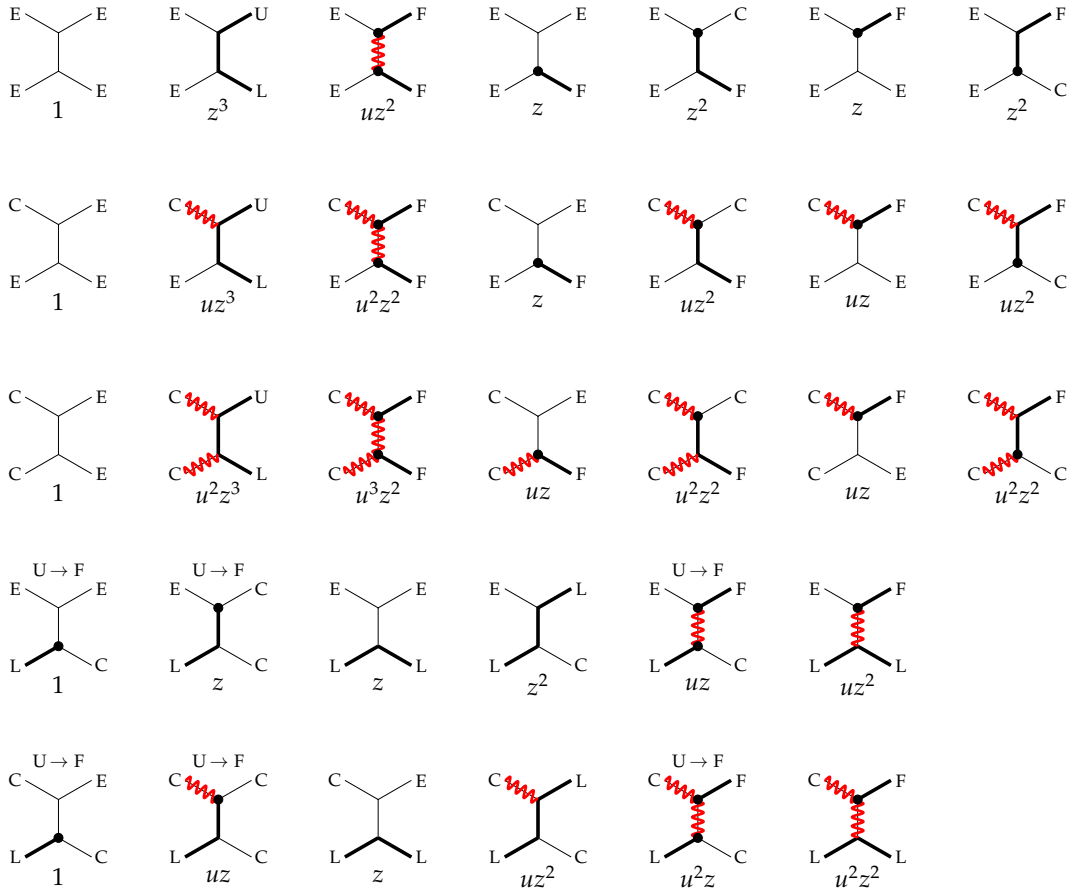

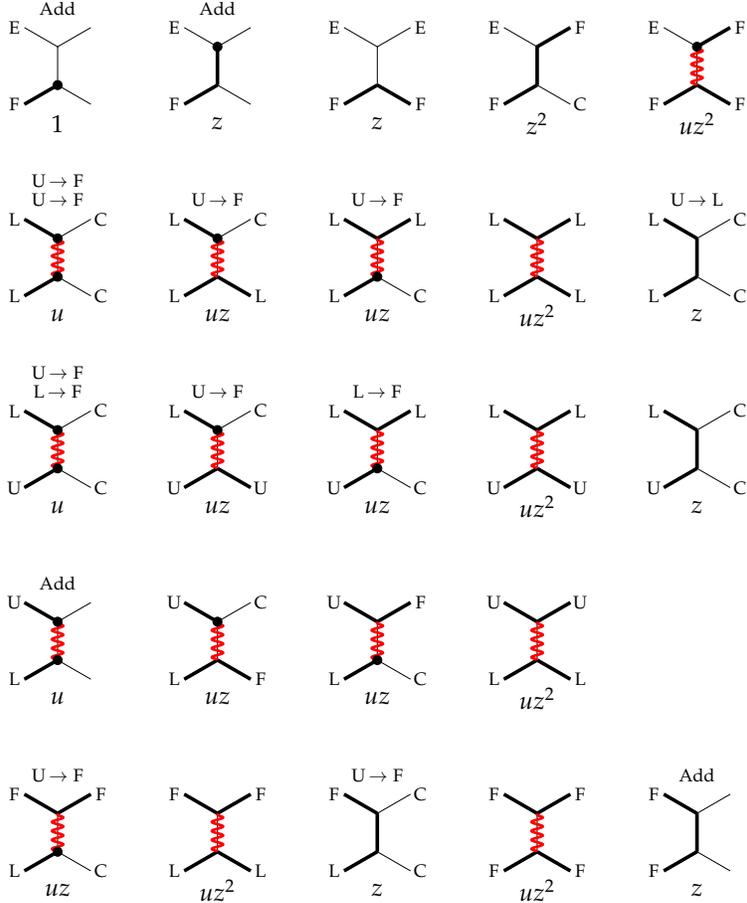
\begin{figure}[ht]
\addtocounter{figure}{-1}
\begin{center}
\scalebox{0.85}{
\begin{tikzpicture}

\scriptsize

\def\xs{0.519615} \def\ys{0.3}

\begin{scope}[yshift=12 cm]
\foreach \x in {0,...,4} {
\draw[xshift= 2.5*\x cm] (-\xs,-\ys)--(0,0)--(\xs,-\ys) (-\xs,3*\ys)--(0,2*\ys)--(\xs,3*\ys) (0,0)--(0,2*\ys);
\draw[xshift= 2.5*\x cm,ultra thick]  (-\xs,-\ys)--(0,0);
\node[xshift= 2.5*\x cm] at (-1.3*\xs,-\ys) {\ef}; \node[xshift= 2.5*\x cm] at (-1.3*\xs,3*\ys) {\ee}; 
}

\node at (0,-2*\ys) {\normalsize $1$}; 
\filldraw (0,0) circle (2pt);  \node at (0,4*\ys) {Add };

\begin{scope}[xshift=2.5 cm]
\draw[ultra thick]  (0,0)--(0,2*\ys);
\filldraw (0,2*\ys) circle (2pt);
\node at (0,-2*\ys) {\normalsize $\ew$};  \node at (0,4*\ys) {Add};
\end{scope}

\begin{scope}[xshift=5 cm]
\draw[ultra thick]  (0,0)--(\xs,-\ys);
\node  at (1.3*\xs,-\ys) {\ef}; \node at (1.3*\xs,3*\ys) {\ee}; \node at (0,-2*\ys) {\normalsize $\ew$};
\end{scope}

\begin{scope}[xshift=7.5 cm]
\draw[ultra thick]  (0,0)--(0,2*\ys)--(\xs,3*\ys);
\node at (1.3*\xs,-\ys) {\ec}; \node at (1.3*\xs,3*\ys) {\ef}; \node at (0,-2*\ys) {\normalsize $\ew^2$};
\end{scope}

\begin{scope}[xshift=10 cm]
\draw[ultra thick]  (0,0)--(\xs,-\ys) (0,2*\ys)--(\xs,3*\ys);
\draw[very thick,red,decorate,decoration=snake,segment length=4] (0,0) -- (0,2*\ys);
\filldraw  (0,2*\ys) circle (2pt);
\node  at (1.3*\xs,-\ys) {\ef}; \node at (1.3*\xs,3*\ys) {\ef}; \node at (0,-2*\ys) {\normalsize $\cw\ew^2$};  
\end{scope}

\end{scope}

\begin{scope}[yshift=9 cm]
\foreach \x in {0,...,4} {
\draw[xshift= 2.5*\x cm] (-\xs,-\ys)--(0,0)--(\xs,-\ys) (-\xs,3*\ys)--(0,2*\ys)--(\xs,3*\ys) (0,0)--(0,2*\ys);
\draw[xshift= 2.5*\x cm] [ultra thick]  (0,0)--(-\xs,-\ys) (0,2*\ys)--(-\xs,3*\ys);
\node[xshift= 2.5*\x cm] at (-1.3*\xs,-\ys) {\el}; \node[xshift= 2.5*\x cm] at (-1.3*\xs,3*\ys) {\el}; 
}

\draw[very thick,red,decorate,decoration=snake,segment length=4] (0,0) -- (0,2*\ys);
\filldraw  (0,0) circle (2pt) (0,2*\ys) circle (2pt);
\node  at (1.3*\xs,-\ys) {\ec}; \node at (1.3*\xs,3*\ys) {\ec}; \node at (0,-2*\ys) {\normalsize $\cw$};  \node at (0,4*\ys) {\eu $\, \to$ \ef }; \node at (0,5*\ys) {\eu $\, \to$ \ef };

\begin{scope}[xshift=2.5 cm]
\draw[ultra thick]  (\xs,-\ys)--(0,0);
\draw[very thick,red,decorate,decoration=snake,segment length=4] (0,0) -- (0,2*\ys);
\filldraw   (0,2*\ys) circle (2pt);
\node  at (1.3*\xs,-\ys) {\el}; \node at (1.3*\xs,3*\ys) {\ec}; \node at (0,-2*\ys) {\normalsize $\cw\ew$};  \node at (0,4*\ys) {\eu $\, \to$ \ef };
\end{scope}

\begin{scope}[xshift=5 cm]
\draw[ultra thick]  (0,2*\ys)--(\xs,3*\ys);
\draw[very thick,red,decorate,decoration=snake,segment length=4] (0,0) -- (0,2*\ys);
\filldraw  (0,0) circle (2pt) ;
\node  at (1.3*\xs,-\ys) {\ec}; \node at (1.3*\xs,3*\ys) {\el}; \node at (0,-2*\ys) {\normalsize $\cw\ew$};  \node at (0,4*\ys) {\eu $\, \to$ \ef };
\end{scope}

\begin{scope}[xshift=7.5 cm]
\draw[ultra thick]  (\xs,-\ys)--(0,0) (0,2*\ys)--(\xs,3*\ys);
\draw[very thick,red,decorate,decoration=snake,segment length=4] (0,0) -- (0,2*\ys);
\node at (1.3*\xs,-\ys) {\el}; \node at (1.3*\xs,3*\ys) {\el}; \node at (0,-2*\ys) {\normalsize $\cw\ew^2$};
\end{scope}

\begin{scope}[xshift=10 cm]
\draw[ultra thick] (0,0)--(0,2*\ys) ;
\node at (1.3*\xs,-\ys) {\ec}; \node at (1.3*\xs,3*\ys) {\ec}; \node at (0,-2*\ys) {\normalsize $\ew$}; \node at (0,4*\ys) {\eu $\, \to$ \el };
\end{scope}

\end{scope}

\begin{scope}[yshift=6 cm]
\foreach \x in {0,...,4} {
\draw[xshift= 2.5*\x cm] (-\xs,-\ys)--(0,0)--(\xs,-\ys) (-\xs,3*\ys)--(0,2*\ys)--(\xs,3*\ys) (0,0)--(0,2*\ys);
\draw[xshift= 2.5*\x cm] [ultra thick]  (0,0)--(-\xs,-\ys) (0,2*\ys)--(-\xs,3*\ys);
\node[xshift= 2.5*\x cm] at (-1.3*\xs,-\ys) {\eu}; \node[xshift= 2.5*\x cm] at (-1.3*\xs,3*\ys) {\el}; 
}

\draw[very thick,red,decorate,decoration=snake,segment length=4] (0,0) -- (0,2*\ys);
\filldraw  (0,0) circle (2pt) (0,2*\ys) circle (2pt);
\node  at (1.3*\xs,-\ys) {\ec}; \node at (1.3*\xs,3*\ys) {\ec}; \node at (0,-2*\ys) {\normalsize $\cw$};  \node at (0,4*\ys) {\el $\, \to$ \ef }; \node at (0,5*\ys) {\eu $\, \to$ \ef };

\begin{scope}[xshift=2.5 cm]
\draw[ultra thick]  (\xs,-\ys)--(0,0);
\draw[very thick,red,decorate,decoration=snake,segment length=4] (0,0) -- (0,2*\ys);
\filldraw   (0,2*\ys) circle (2pt);
\node  at (1.3*\xs,-\ys) {\eu}; \node at (1.3*\xs,3*\ys) {\ec}; \node at (0,-2*\ys) {\normalsize $\cw\ew$};  \node at (0,4*\ys) {\eu $\, \to$ \ef };
\end{scope}

\begin{scope}[xshift=5 cm]
\draw[ultra thick]  (0,2*\ys)--(\xs,3*\ys);
\draw[very thick,red,decorate,decoration=snake,segment length=4] (0,0) -- (0,2*\ys);
\filldraw  (0,0) circle (2pt) ;
\node  at (1.3*\xs,-\ys) {\ec}; \node at (1.3*\xs,3*\ys) {\el}; \node at (0,-2*\ys) {\normalsize $\cw\ew$};  \node at (0,4*\ys) {\el $\, \to$ \ef };
\end{scope}

\begin{scope}[xshift=7.5 cm]
\draw[ultra thick]  (\xs,-\ys)--(0,0) (0,2*\ys)--(\xs,3*\ys);
\draw[very thick,red,decorate,decoration=snake,segment length=4] (0,0) -- (0,2*\ys);
\node at (1.3*\xs,-\ys) {\eu}; \node at (1.3*\xs,3*\ys) {\el}; \node at (0,-2*\ys) {\normalsize $\cw\ew^2$};
\end{scope}

\begin{scope}[xshift=10 cm]
\draw[ultra thick] (0,0)--(0,2*\ys) ;
\node at (1.3*\xs,-\ys) {\ec}; \node at (1.3*\xs,3*\ys) {\ec}; \node at (0,-2*\ys) {\normalsize $\ew$}; 
\end{scope}

\end{scope}

\begin{scope}[yshift=3 cm]
\foreach \x in {0,...,3} {
\draw[xshift= 2.5*\x cm] (-\xs,-\ys)--(0,0)--(\xs,-\ys) (-\xs,3*\ys)--(0,2*\ys)--(\xs,3*\ys) (0,0)--(0,2*\ys);
\draw[xshift= 2.5*\x cm] [ultra thick]  (0,0)--(-\xs,-\ys) (0,2*\ys)--(-\xs,3*\ys);
\node[xshift= 2.5*\x cm] at (-1.3*\xs,-\ys) {\el}; \node[xshift= 2.5*\x cm] at (-1.3*\xs,3*\ys) {\eu}; 
}

\draw[very thick,red,decorate,decoration=snake,segment length=4] (0,0) -- (0,2*\ys);
\filldraw  (0,0) circle (2pt) (0,2*\ys) circle (2pt);
\node at (0,-2*\ys) {\normalsize $\cw$};  \node at (0,4*\ys) {Add };  

\begin{scope}[xshift=2.5 cm]
\draw[ultra thick]  (\xs,-\ys)--(0,0);
\draw[very thick,red,decorate,decoration=snake,segment length=4] (0,0) -- (0,2*\ys);
\filldraw   (0,2*\ys) circle (2pt);
\node  at (1.3*\xs,-\ys) {\ef}; \node at (1.3*\xs,3*\ys) {\ec}; \node at (0,-2*\ys) {\normalsize $\cw\ew$};  
\end{scope}

\begin{scope}[xshift=5 cm]
\draw[ultra thick]  (0,2*\ys)--(\xs,3*\ys);
\draw[very thick,red,decorate,decoration=snake,segment length=4] (0,0) -- (0,2*\ys);
\filldraw  (0,0) circle (2pt) ;
\node  at (1.3*\xs,-\ys) {\ec}; \node at (1.3*\xs,3*\ys) {\ef}; \node at (0,-2*\ys) {\normalsize $\cw\ew$};  
\end{scope}

\begin{scope}[xshift=7.5 cm]
\draw[ultra thick]  (\xs,-\ys)--(0,0) (0,2*\ys)--(\xs,3*\ys);
\draw[very thick,red,decorate,decoration=snake,segment length=4] (0,0) -- (0,2*\ys);
\node at (1.3*\xs,-\ys) {\el}; \node at (1.3*\xs,3*\ys) {\eu}; \node at (0,-2*\ys) {\normalsize $\cw\ew^2$};
\end{scope}
 
\end{scope}

\begin{scope}[yshift=0 cm]
\foreach \x in {0,...,2} {
\draw[xshift= 2.5*\x cm] (-\xs,-\ys)--(0,0)--(\xs,-\ys) (-\xs,3*\ys)--(0,2*\ys)--(\xs,3*\ys) (0,0)--(0,2*\ys);
\draw[xshift= 2.5*\x cm] [ultra thick]  (0,0)--(-\xs,-\ys) (0,2*\ys)--(-\xs,3*\ys);
\node[xshift= 2.5*\x cm] at (-1.3*\xs,-\ys) {\el}; \node[xshift= 2.5*\x cm] at (-1.3*\xs,3*\ys) {\ef}; 
}

\begin{scope}[xshift=0 cm]
\draw[ultra thick]  (0,2*\ys)--(\xs,3*\ys);
\draw[very thick,red,decorate,decoration=snake,segment length=4] (0,0) -- (0,2*\ys);
\filldraw  (0,0) circle (2pt) ;
\node  at (1.3*\xs,-\ys) {\ec}; \node at (1.3*\xs,3*\ys) {\ef}; \node at (0,-2*\ys) {\normalsize $\cw\ew$};  \node at (0,4*\ys) {\eu $\, \to$ \ef };
\end{scope}

\begin{scope}[xshift=2.5 cm]
\draw[ultra thick]  (\xs,-\ys)--(0,0) (0,2*\ys)--(\xs,3*\ys);
\draw[very thick,red,decorate,decoration=snake,segment length=4] (0,0) -- (0,2*\ys);
\node at (1.3*\xs,-\ys) {\el}; \node at (1.3*\xs,3*\ys) {\ef}; \node at (0,-2*\ys) {\normalsize $\cw\ew^2$};
\end{scope}

\begin{scope}[xshift=5 cm]
\draw[ultra thick] (0,0) -- (0,2*\ys);
\node at (1.3*\xs,-\ys) {\ec}; \node at (1.3*\xs,3*\ys) {\ec}; \node at (0,-2*\ys) {\normalsize $\ew$}; \node at (0,4*\ys) {\eu $\, \to$ \ef };
\end{scope}

\end{scope}

\begin{scope}[xshift=7.5cm,yshift=0 cm]
\foreach \x in {0,...,1} {
\draw[xshift= 2.5*\x cm] (-\xs,-\ys)--(0,0)--(\xs,-\ys) (-\xs,3*\ys)--(0,2*\ys)--(\xs,3*\ys) (0,0)--(0,2*\ys);
\draw[xshift= 2.5*\x cm] [ultra thick]  (0,0)--(-\xs,-\ys) (0,2*\ys)--(-\xs,3*\ys);
\node[xshift= 2.5*\x cm] at (-1.3*\xs,-\ys) {\ef}; \node[xshift= 2.5*\x cm] at (-1.3*\xs,3*\ys) {\ef}; 
}

\begin{scope}[xshift=0 cm]
\draw[ultra thick]  (\xs,-\ys)--(0,0) (0,2*\ys)--(\xs,3*\ys);
\draw[very thick,red,decorate,decoration=snake,segment length=4] (0,0) -- (0,2*\ys);
\node at (1.3*\xs,-\ys) {\ef}; \node at (1.3*\xs,3*\ys) {\ef}; \node at (0,-2*\ys) {\normalsize $\cw\ew^2$};
\end{scope}

\begin{scope}[xshift=2.5 cm]
\draw[ultra thick] (0,0) -- (0,2*\ys);
 \node at (0,-2*\ys) {\normalsize $\ew$}; \node at (0,4*\ys) {Add };
\end{scope}

\end{scope}

\end{tikzpicture}  
}
\end{center}
\caption{(continued) Updating rules for the TM enumeration of honeycomb ISAW.
}
\end{figure}

Below we describe how some of these rules are derived:

\begin{description}

\item{\ee\ee:} The bottom and top left edges are both empty. We have five possible output states. 

We can 
leave the bottom and right edges empty (\ee\ee) which means no new edges or contacts were added
so the weight is just 1. 

We can insert a new partial loop (\el\eu) thus adding three new edges with weight $z^3$.

We can add two new free edges on the right (\ef\ef) (this is only allowed provided the source has 
no free ends). This creates a new contact along the vertical edge so the weight is $\cw\ew^2$.

Finally we can add a free edge on the bottom (\ef\ee) or top edge (\ee\ef) on the right (allowed
provided there is at most one free edge in the source). In this case we may or may not occupy
the vertical edge as well so there are two instances with weights $\ew$ and $\ew^2$, respectively.
 
\item{\el\ee:} The bottom left edge has a lower loop-edge while
the top left edge is empty.  

Firstly, we may terminate the lower-loop edge. This creates a new
degree-1 vertex so it is only allowed if there is at most 1 free edge in the source. 
Here we may or may not occupy the vertical edge. We also have to relabel the matching upper loop-edge of the 
now discontinued lower loop-edge as a free edge. Such an edge relabelling is indicated above the
configuration (in this case by $\eu \to \ef$). 

Secondly, we can continue the loop edge along the bottom right edge (weight $\ew$) or along both vertical and top right edges (weight $\ew^2$) .
 
Thirdly, we can terminate the lower-loop edge (relabelling) and add a new free edge on the top right edge (this is only allowed provided the source has 
no free edges). We insert a single edge and create a contact along the vertical edge so the weight is $\cw\ew$.

Finally, we continue the loop-edge along the bottom and insert a new free edge on the top while creating a 
contact along the vertical edge so the weight is $\cw\ew^2$.

\item{\ef\ee:} The bottom left edge is a free edge.

We can leave both right edges empty. This creates a separate component and
is only allowed if the resulting graph is a valid ISAW. That is,
the source contains no other occupied edges (and if required both
the bottom and top of the rectangle has been touched). The partial
generating function is added to the running total for the ISAW generating function. We mark this
possibility by an `Add' above the configuration.

We can continue the free end along the bottom edge or the vertical and top edges.

We continue the free end along the bottom edge and insert a new free edge on top.

\item{\el\el} One or both lower edges can terminate or both can be continued with the by
now obvious results. Finally, we can occupy the vertical edge which means that
we join two loops together. The matching upper loop-edge of the top most lower
edge becomes the lower edge of the joined loop.  Such an edge relabelling is indicated  by $\eu \to \el$. 

\item{\el\eu:} The two loop-edges belong to the same loop so terminating both edges results in
a separate component so we check and add if permitted. Otherwise, both or at least one
edge is continued to the right.  
 
\end{description}

\subsubsection{Calculation of the ISAW generating function}

To calculate the ISAW series we use our TM algorithm to count the number of
ISAW in rectangles of size $W \times L$ using the TM traversal direction 1 of
the left panel of Figure~\ref{fig:transfer} when $L\geq W$ and the traversal direction 2 of
the right panel when $W<L$. We calculated the series for walks up to 75 steps.
This required calculations for rectangles up to $W=26$  in direction 1. 
For a given width all required lengths can be obtained in a single calculation
(similarly for direction 2). The final series is then obtained by simply adding up
the results for all rectangles.

The ISAW generating function is a two variable function $\CGf(\cw,\ew)$. The coefficient
of $\ew^k$ is a polynomial in $\cw$. 
In order to calculate a series expansion for $\CGf(\cw,\ew)$ we would generally
need to store a two variable polynomial (truncating the series at order $n=75$ in $\ew$).
The degree of the polynomial of $z^{75}$ is 28. Such a calculation
requires a lot of memory and it makes it more difficult to write an efficient parallel
version of the TM algorithm  (for details on parallelisation see \cite{IJ04,IJ06,PolygonBook}). We have therefore
taken a different approach. Since the coefficient of $\ew^k$ is a polynomial in $\cw$
we can reconstruct the coefficients of this polynomial from evaluations at 
integer values of $\cw$. We calculate the series for $\CGf(\cw=i,\ew), i\in[0,28]$ 
and then use these evaluated series to reconstruct the actual series for $\CGf(\cw,\ew)$.
Thankfully we do not generally have to use 29 evaluations. The number of contacts
reaches the maximum of $28$ at width 8 and then decreases as the width of the rectangle increases. 
In this way our calculations require less memory and we get a more stable parallel algorithm, but 
everything in life has a cost and in this case the need for several evaluations at integer points 
means we had to use more CPU time. The typical trade-off was that we saved at least
an order of magnitude in memory but had to use about 4 times as much CPU time.
 
The integer coefficients of the series expansions become very large and 
in order to handle this the calculations were performed using modular
arithmetic. The series were calculated modulo prime number $p_i$
and the coefficients reconstructed at the end using the Chinese remainder theorem.
The equations  used to determine the polynomial coefficients of $\ew^k$
can readily be solved mod $p_i$, so we did this prior to applying the Chinese remainder theorem.
The algorithm for the calculation of metric properties requires integer multiplication 
so we used the largest prime numbers smaller than $2^{30}$.
In this case we needed 3 primes to represent the integer coefficients.
 
\subsection{Square lattice}

We calculated series for the ISAW generating function $\CGf(\cw,\ew)$ for walks up to 59 steps and for the metric quantities $\ave{R^2_x}_n$ up to 55 steps. 
The algorithm used is a generalisation of the one used by Jensen
to enumerate non-interacting SAW \cite{IJ13} to 79 steps. The changes required to enumerate ISAW are identical to those described above for the honeycomb lattice. 
The updating rules in the square case are simpler since only a single vertex is added when moving the kink in the boundary and furthermore the symmetry of the square lattice means that rectangles of size $W\times L$ and $L\times W$ have the partial generating functions so only one direction of the TM algorithm is required.
However, for square ISAW the number of interactions increase faster with the length of the walk so that for length 59 there is a maximum of 45 interactions.

\section{Results}\label{sec:results}

\subsection{Non-interacting SAWs}\label{ssec:non_interact}

\subsubsection{Honeycomb lattice}

The metric properties of non-interacting honeycomb lattice SAWs have been extensively studied by Jensen in \cite{IJ06}, based on enumerations of SAWs to length 105 steps, and their metric properties to length 96 steps. Writing
\[ \ave{\Res}_n\sim \Me n^{2\nu}, \quad \ave{\Rgs}_n\sim \Mg n^{2\nu}, \quad \ave{\Rms}_n\sim \Mm n^{2\nu},\]
Jensen estimated $\Me=0.8857(1),$ $\Mg=0.12424(4),$ and $\Mm=0.3894(1).$ The quantity $A_\infty$ in (\ref{FN}) is just $\Mg/\Me,$ and $B_\infty$ is $\Mm/\Me.$ Jensen estimated these directly as $0.1403001(2)$ and $0.439635(1)$ respectively. As a consequence, the estimate of $\lim_{n \to \infty} F_n =0.000003(13).$

Jensen also studied the SAW generating function, and pointed out that as well as the dominant singularity at $\ew=\ew_c=1/\sqrt{2+\sqrt{2}},$ with exponent $\gamma=\frac{43}{32}$, there is a singularity on the negative real axis at $\ew=-\ew_c,$ with exponent $-\frac12$, and a conjugate-pair of singularities on the imaginary axes at $\ew=\ew_c^-\approx \pm0.64215i.$ This singularity structure induces two distinct sets of parity effects, meaning that one really needs to look at every fourth term to focus clearly on the dominant singularity. This remark applies equally to the case of interacting walks, considered below.
\subsubsection{Square lattice}

For square-lattice non-interacting SAWs, a corresponding series analysis, as well as an exhaustive Monte Carlo analysis is given in \cite{CGJPRS} by Caracciolo et al. They estimated the corresponding amplitudes for square-lattice SAWs (also triangular lattice SAWs, but we will not discuss those here) as $\Me=0.77121(4),$ $\Mg=0.108207(7)$ and $\Mm=0.33903(4).$ They also estimated the ratios directly, and found $A_\infty=0.140296(6),$ and $B_\infty=0.439649(9).$ These give $\lim_{n \to \infty} F_n =0.000036(37).$ It is clear that the amplitude ratios $A_\infty$ and $B_\infty$ appear to be lattice-independent, as is expected.

In the case of the square lattice, as well as the physical singularity in the SAW generating function at $\ew=\ew_c\approx0.379052277755161,$ \cite{JSG16} with exponent $\gamma=\frac{43}{32}$, there is another singularity at $\ew=-\ew_c,$ with exponent $-\frac12$. This induces a parity effect on numerical properties, such as ratios of successive terms, and so it is customary to look at series comprised of every second term to eliminate the effect of the singularity on the negative axis.

\subsection{Interacting SAWs}\label{ssec:interact}

As first conjectured, with supporting numerical evidence based on series expansions, in \cite{OPBG}, and subsequently supported by rather precise Monte Carlo analysis in \cite{CGPP}, the quantity 
\begin{equation}\label{FN2}
    F_n=\frac{23}{8}A_n-2B_n+\frac{1}{2}
\end{equation}
should vanish at the $\theta$ temperature as $n \to \infty$. 

More remarkably, as proved in \cite{CGJPRS} for the non-interacting case, so should $$\lim_{n \to \infty} n F_n =0.$$ As a consequence, we expect 
\[F_n = \bigO(n^{-3/2}).\]

\subsubsection{Honeycomb lattice}\label{sssec:interact_hc}

Using the series expansions derived as described in Section~\ref{sec:series}, we calculated $\langle \Res\rangle_n,$ $\langle \Rgs\rangle_n,$ and $\langle \Rms\rangle_n,$ as a function of $\cw=\exp(-\epsilon/kT)$, and calculated the amplitude ratios $A_n(\cw)$ and $B_n(\cw)$, and thus the sequence $\{F_n(\cw)\},$ for a range of $\cw$ values. Extrapolating $F_n(\cw)$ against $n^{-3/2},$ and seeking that value of $\cw$ which resulted in the $n \to \infty$ limit of $F_n(\cw)$ to vanish, we quickly found the critical value $\cw=\cwc$ to be around $\cw=2.77,$ and more careful analysis allowed us to refine this estimate to $2.767 \pm 0.002.$ The only other estimate of this quantity we can find in the literature is an old (1989) calculation of Poole et al.~\cite{PCJS} who give $\cwc=2.69 \pm 0.14,$ some two orders of magnitude less precise than our estimate.

We also calculated the individual amplitudes directly at this value of $\cwc$, in the simplest possible way, by just extrapolating the sequence $\langle R^2 \rangle_n/n^{2\nu}$  against $\frac1n$. In the notation of the previous subsection, we estimated $\Me = 1.525(2),$ $\Mg=0.2772(3)$ and $\Mm=0.7795(5).$ This gives for the amplitude ratios $A_\infty=0.1817(5)$ and $B_\infty=0.5111(10).$ From (\ref{FN2}) we find $F=0.00019(340).$ Greater precision could undoubtedly be obtained by a more sophisticated analysis, such as that performed in \cite{CGJPRS}, but as the precise values of these amplitudes are of little interest, we have not carried out any further analysis. We remark that the amplitude ratio $A_\infty$ was first estimated by Chang et al.~\cite{CSM90} by the scanning simulation method. Their estimate $A_\infty = 0.179 \pm 0.003$ is in agreement with our more precise estimate. More precise estimates were given in \cite{CGPP}, albeit for the square lattice (though these amplitude ratios are expected to be universal). They give $A_\infty=0.18151(10)$ and $B_\infty=0.51106(31).$ These are in complete agreement with our series results.

\begin{figure}[ht]
\centering
\begin{tikzpicture}
\node at (0,0) [anchor=south west] {\includegraphics[width=0.5\textwidth]{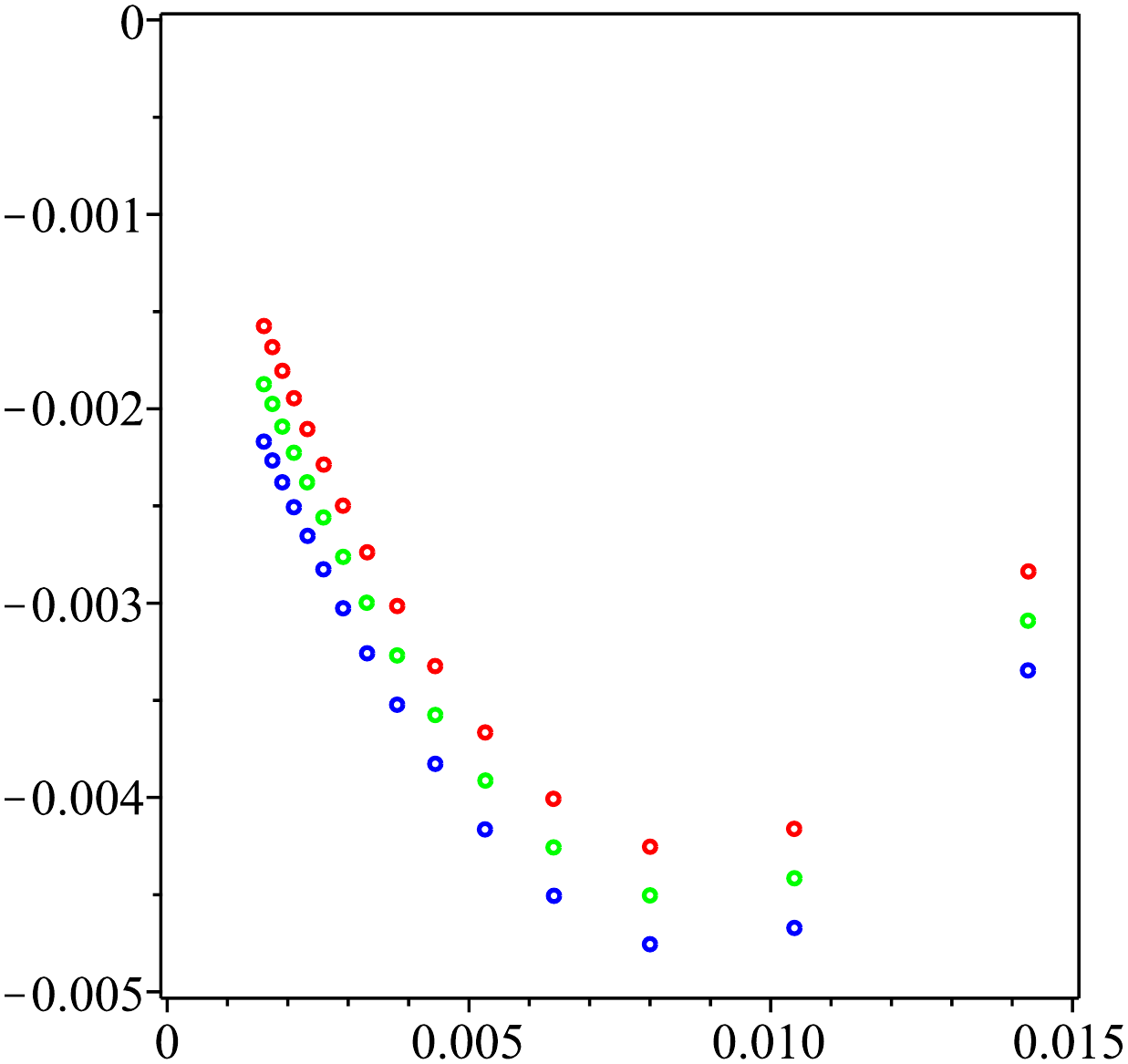}};
\node at (-0.2,4.3) {$F_n$};
\node at (4.7,-0.2) {$n^{-3/2}$};
\end{tikzpicture}
 \caption{Honeycomb lattice. A plot of $F_n=\frac{23}{8}A_n-2B_n+\frac{1}{2}$ against $n^{-3/2}$ for $\cwc=2.777,\,\,2.767,\,\,2.757$ reading from top to bottom. We claim that the central curve appears to be headed toward the origin.} 

 \label{fig:honeyfn}
\end{figure}

One typical piece of numerical evidence for our estimate of $\cwc$ is shown in Figure \ref{fig:honeyfn} where the values of $F_n$ are extrapolated against $n^{-3/2}$ for a range of $\cw$ values. While some curvature in the plots is evident, we argue that the central curve, corresponding to $\cwc=2.767$ appears to pass through the origin, while the plot corresponding to $\cwc=2.777$ extrapolates to a value above zero, and the plot corresponding to $\cwc=2.757$ extrapolates to a value below zero.

It is significant that the curve passes through a minimum around $n=25.$ If we had many fewer terms, it would not be at all obvious that $F_n$ was approaching 0. This will turn out to be significant in our analysis, below, of the corresponding square-lattice data.

A completely different way to estimate $\cwc$ is to calculate the critical exponent $\nu$ of any of the three metric properties that we have studied. For $\cw < \cwc$ we expect $\nu = \frac34,$ which should change discontinuously to $\nu=\frac47$ at $\cw=\cwc$ and to $\nu = \frac12$ for $\cw > \cwc.$ As we are dealing with finite series, we will not see discontinuities in the exponent estimates, but we can expect that at $\cw=\cwc,$ the exponent $\nu$ should be $\frac47$. Thus our other approach to estimating $\cwc$ is to vary $\cw$ and analyse the resulting metric series for their associated exponents.

This we do as follows: Let $\langle R^2 \rangle_n \sim An^{2\nu}$ for some generic length metric. Then $$r_n \equiv \frac{\langle R^2 \rangle_n}{\langle R^2 \rangle_{n-1}} \sim 1+\frac{2\nu}{n}.$$ So one can estimate $\nu$ by studying the sequence $\nu_n = (r_n-1)\frac{n}{2}.$
Extrapolating $\nu_n$ against $\frac1n$ should give an estimate of $\nu$ as $n \to \infty.$ If there are no confluent singular terms, the plot should be linear. Otherwise there will be some curvature, but the plot is still able to be extrapolated.

Because of other singularities on the negative real axis and the imaginary axes in the case of the honeycomb lattice, to eliminate oscillations produced by these singularities, we in practice calculate 
\[r_n = \frac{\langle R^2 \rangle_n}{\langle R^2 \rangle_{n-4}} \sim 1+\frac{8\nu}{n},\]
and estimate $\nu$ from the sequence $\nu_n = (r_n-1)\frac{n}{8}.$

\begin{figure}[ht]
\centering
\begin{tikzpicture}
\node at (0,0) [anchor=south west] {\includegraphics[width=0.5\textwidth]{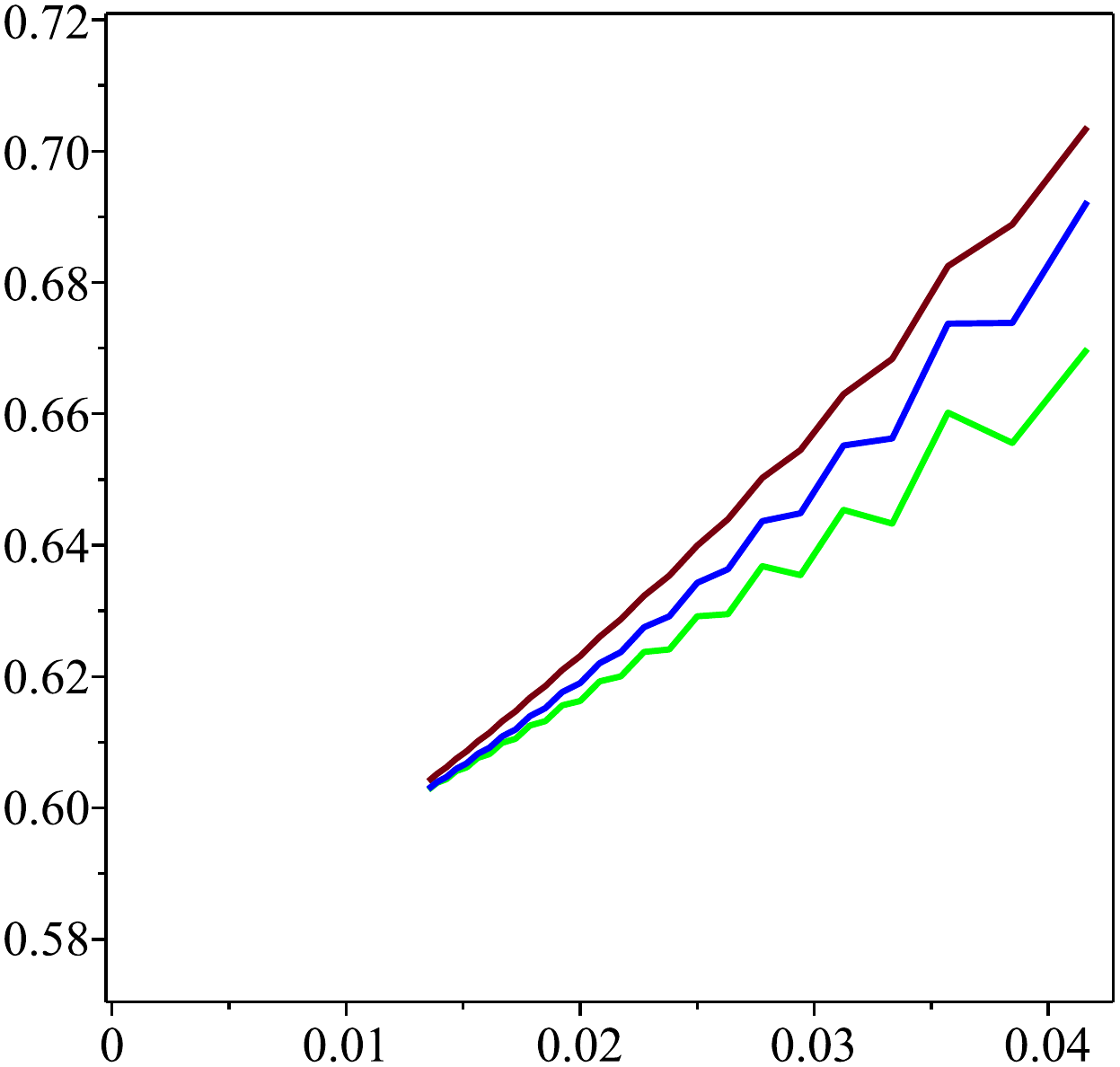}};
\node at (-0.2,4.4) {$\nu_n$};
\node at (4.5,-0.3) {$\frac1n$};
\end{tikzpicture}
 \caption{Honeycomb lattice. Estimates of the exponent $\nu$ at $\cw=2.767$ from $\langle \Rgs\rangle_n,\,\, \langle \Rms\rangle_n,\,\, \langle \Res\rangle_n$ reading from top to bottom. We claim that all three curves appear to be heading toward the origin, corresponding to $\nu=\frac47$.}
 \label{fig:honeynu}
\end{figure}

In Figure \ref{fig:honeynu} we show the plots for the three metric properties at our central estimate of $\cwc.$ It can be seen that all are plausibly approaching $\frac47$. The top curve, (in red, for those viewing in colour) is $\langle \Rgs\rangle_n,$ the middle (blue) curve is the data for $ \langle \Rms\rangle_n,$ and the bottom (green) curve is the data for $ \langle \Res\rangle_n.$

Finally, we repeated the method of analysis introduced in \cite{OPBG}. From  (\ref{FN2}) we see that 
\[\lim_{n \to \infty} G_n =\frac{4B_\infty-1}{2A_\infty}=\frac{23}{8}.\]
We then define 
\[H_n = \prod_{m=0}^n G_m,\]
and note that 
\[\sum H_n \ew^n \sim c(1-G_\infty \ew)^\lambda.\]
Then a simple ratio analysis allows one to estimate $G_\infty,$ which should equal $\frac{23}{8}$ at the $\theta$-point. (The value of the exponent $\lambda$ is irrelevant to this calculation). 

\begin{figure}[ht]
\centering
\begin{tikzpicture}
\node at (0,0) [anchor=south west] {\includegraphics[width=0.5\textwidth]{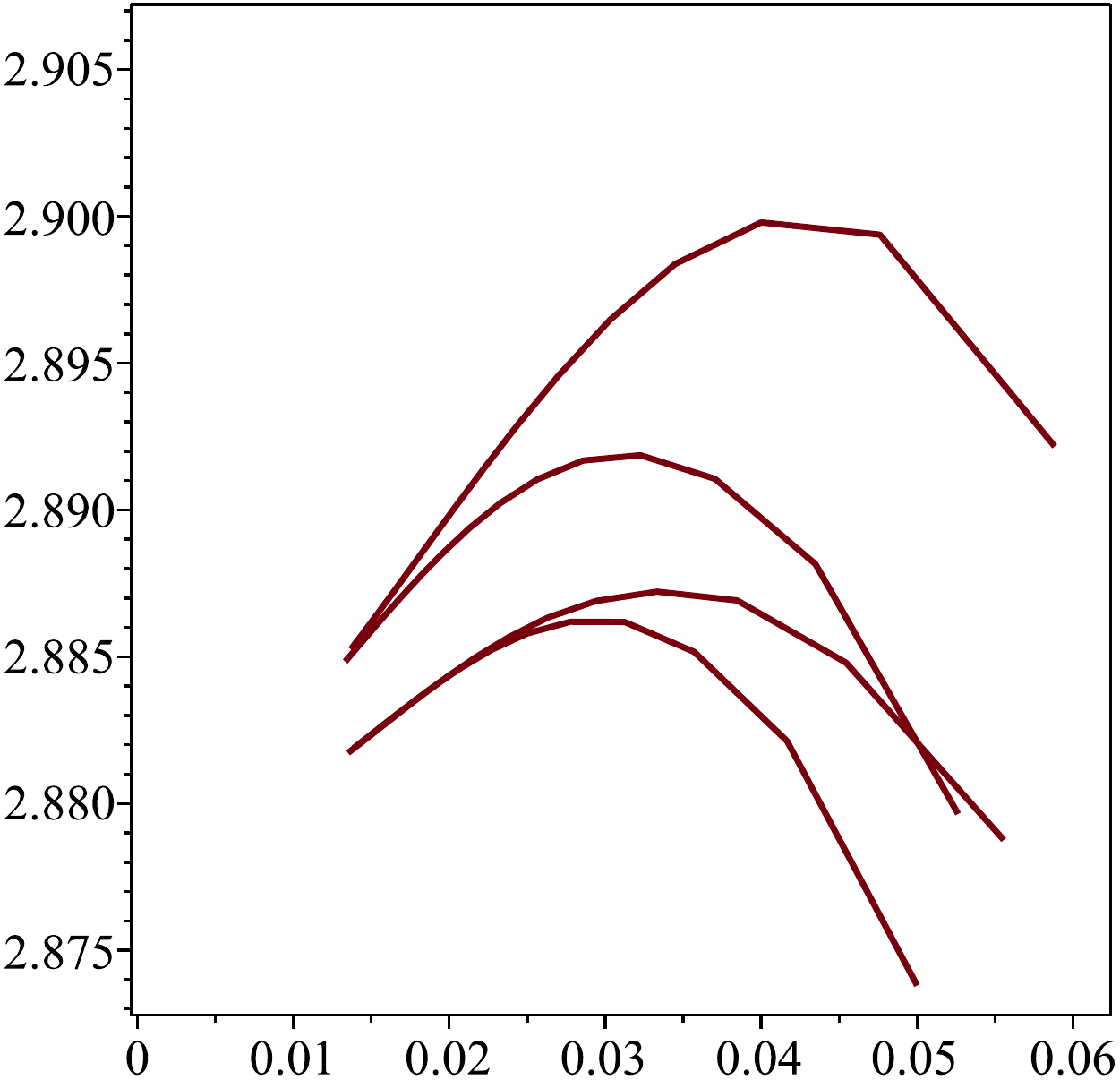}};
\node at (-0.2,4.4) {$r_n$};
\node at (4.5,-0.3) {$\frac1n$};
\end{tikzpicture}
 \caption{Honeycomb lattice. Plot of $r_n=H_n/H_{n-4}$ against $\frac{1}{n}=1/(4j+k)$ for $k=0,1,2,3$. All curves should extrapolate to $\frac{23}{8}$ as $n \to \infty.$}
 \label{fig:honeyG}
\end{figure}

In Figure \ref{fig:honeyG} we show plots of the ratios $H_n/H_{n-1}$ against $\frac1n$. Again, because of the singularities on the negative real and imaginary axes, this plot has a 4-term periodicity. We have therefore broken the data into four sets, according as $n$~(mod 4) is $0,1,2,3.$ This gives four distinct plots, all of which should go to the same limit, and that limit should be $\frac{23}{8}=2.875.$ As can be seen, the plots behave in exactly the expected way. This test is moderately sensitive. If the estimate of $u_c$ is increased or decreased by $0.005$ the corresponding plots can be seen to go to a limit above or below $\frac{23}{8}$ respectively.

We also studied the ISAW series at $u=u_c,$ by the method of differential approximants. Just as in the non-interacting case discussed above, we found four singularities. The dominant singularity at $\ew=\ew_c \approx 0.481846,$ with an exponent of about $1.11,$ compared to the predicted value of $\frac87=1.1428\ldots,$ and a second singularity at $\ew=-\ew_c$ with a small exponent that is difficult to estimate, but appears to be about $-0.1 \pm 0.1,$ and a conjugate pair of singularities at $\ew=\ew_c^- \approx \pm 0.507$ with an exponent we cannot estimate.

\subsubsection{Square lattice}\label{sssec:interact_sq}

For the square-lattice, we have fewer terms than for the honeycomb. One consequence of this is that the corresponding plot of $F_n$ against $n^{-3/2},$ (shown in Figure \ref{fig:honeyfn} for the honeycomb lattice), appears to be approaching a minimum, but we have insufficient terms to see this clearly, and, more importantly, to extrapolate the plot after it has passed through a minimum. So we turn instead to the other two methods used above for the analysis.

In earlier work \cite{OPBG} based on shorter series, analysis of the metric properties to estimate $\cwc$ gave rise to the estimate $\log(\cwc)=0.665(5).$ A more recent Monte Carlo analysis in \cite{CGPP} using walks up to 3200 steps, gave rise to the estimate $\log(\cwc)=0.6673(5).$ Our analysis suggests that this is slightly high, our result being $\log(\cwc) = 0.6665(5),$ corresponding to $\cwc=1.9474(10).$

\begin{figure}[ht]
\centering
\begin{tikzpicture}
\node at (0,0) [anchor=south west] {\includegraphics[width=0.5\textwidth]{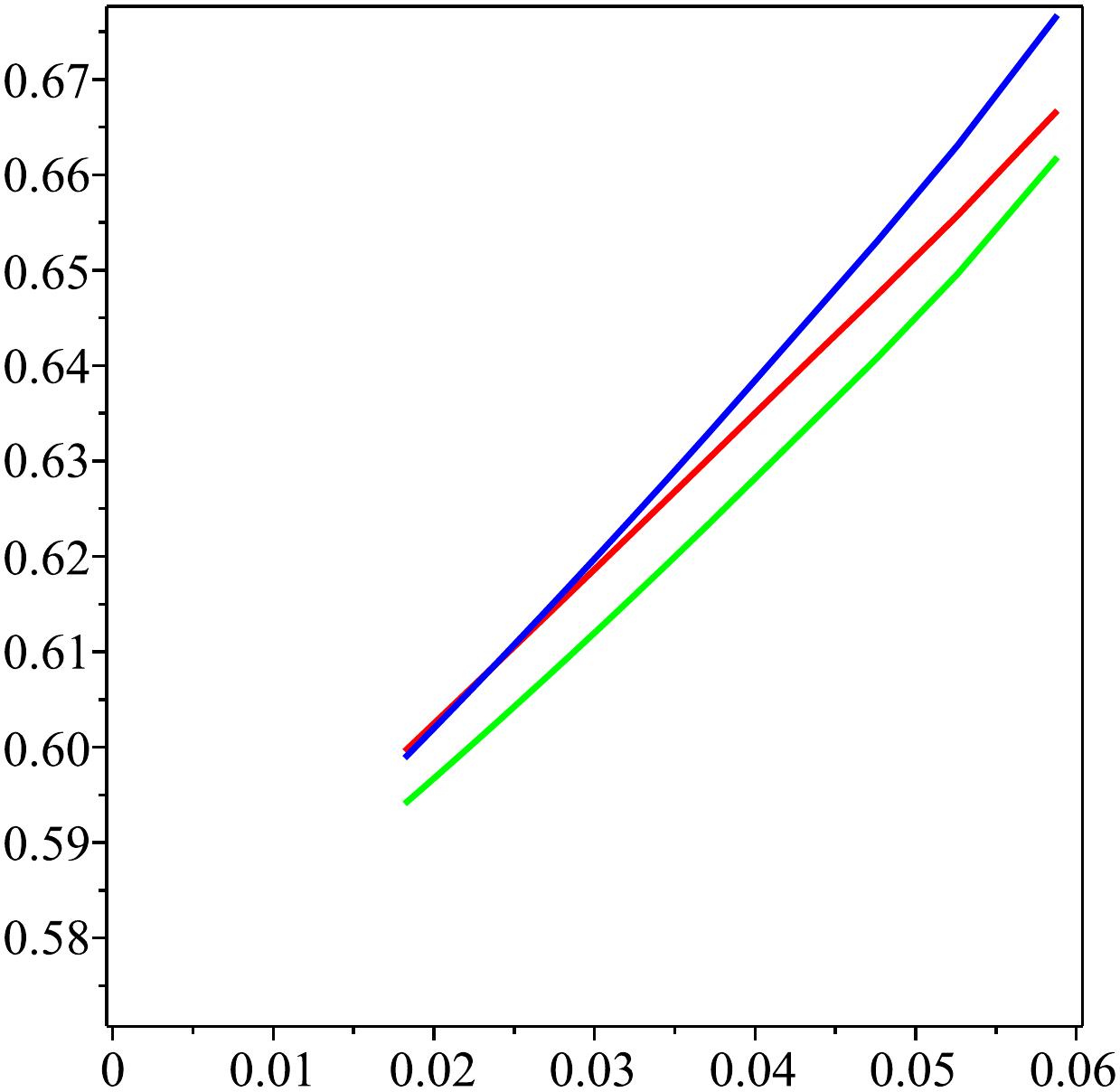}};
\node at (-0.2,4.4) {$\nu_n$};
\node at (4.5,-0.3) {$\frac1n$};
\end{tikzpicture}
 \caption{Square lattice. Estimates of the exponent $\nu$ at $\cw=1.9474$ from $\langle \Rms\rangle_n,\,\, \langle \Rgs\rangle_n,\,\, \langle \Res\rangle_n$ reading from top to bottom. We claim that all three curves appear to be heading toward the origin, corresponding to $\nu=\frac47$.}
 \label{fig:sqnu}
\end{figure}

In Figure \ref{fig:sqnu} we show the plots for the three metric properties at our central estimate of $\cwc.$ It can be seen that all are plausibly going to the origin, corresponding to $\nu = \frac47$. The top curve, (in blue, for those viewing in colour) is $\langle \Rms\rangle_n,$ the middle (red) curve is the data for $ \langle \Rgs\rangle_n,$ and the bottom (green) curve is the data for $ \langle \Res\rangle_n.$

Finally, we repeated the method of analysis introduced in \cite{OPBG}, as we did for the honeycomb lattice.
However, again we see the effect of a shorter series.

\begin{figure}[ht]
\centering
\begin{tikzpicture}
\node at (0,0) [anchor=south west] {\includegraphics[width=0.5\textwidth]{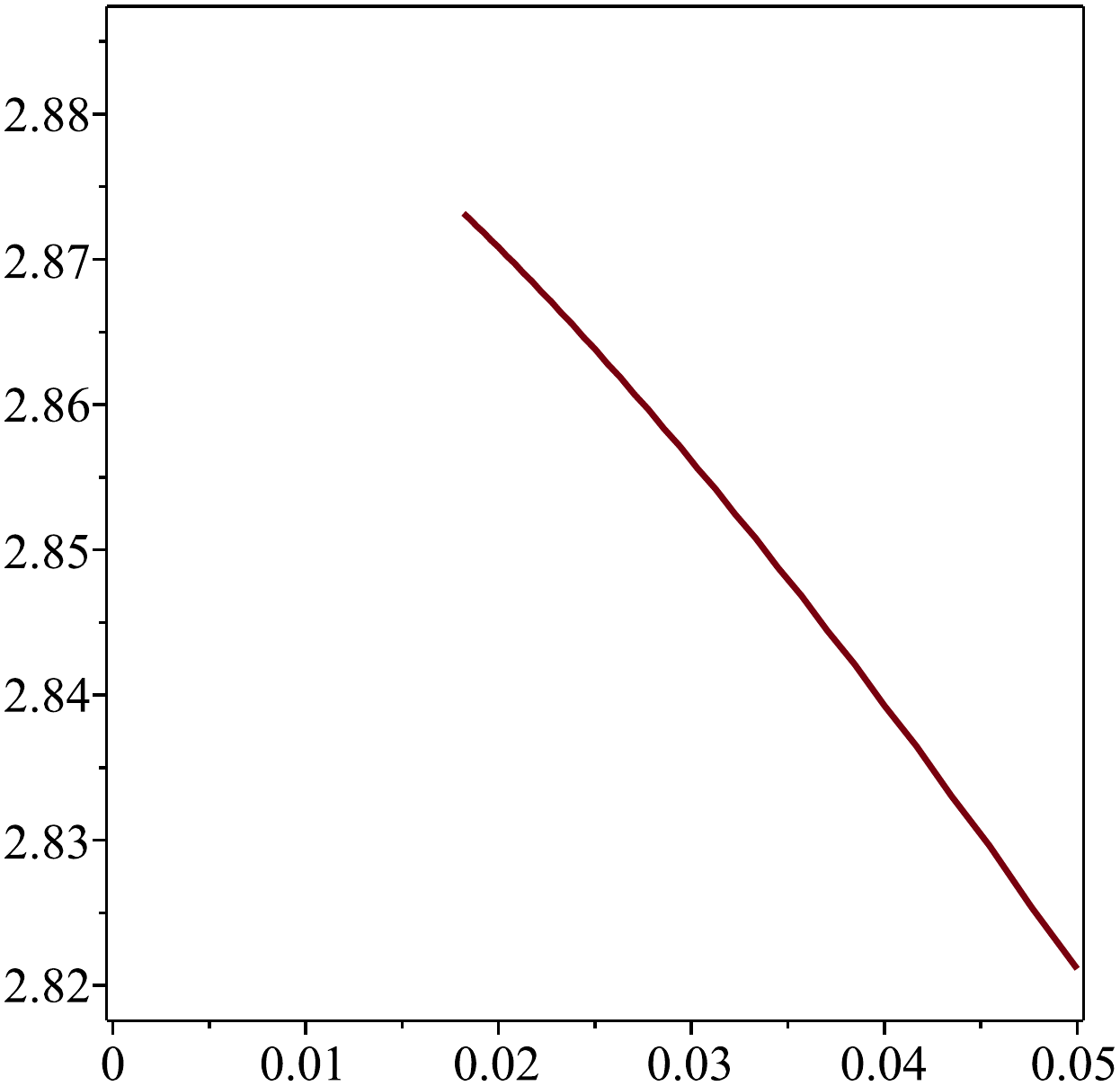}};
\node at (-0.2,4.4) {$r_n$};
\node at (4.5,-0.3) {$\frac1n$};
\end{tikzpicture}

 \caption{Square lattice. Plot of $r_n=\sqrt{H_n/H_{n-2}}$ against $\frac1n$. The curve should extrapolate to $\frac{23}{8}$ as $n \to \infty.$}
 \label{fig:sqG}
\end{figure}

In Figure \ref{fig:sqG} we show a plot of the ratios $\sqrt{H_n/H_{n-2}}$ against $\frac1n.$ Because of singularities on the negative real axis, we have taken ratios of alternate terms. It can be seen that the plot appears to be going to a limit greater than $\frac{23}{8}.$ We believe that this is a short series effect, and that with further terms this plot would pass through a maximum, and then decrease to a value around $\frac{23}{8},$ just like the corresponding plot for the honeycomb lattice.

So we base our analysis for the square lattice just on the metric property analysis, and conclude that $\log(\cwc)=0.6665(5),$ which overlaps the estimate from the best Monte Carlo analysis \cite{CGPP}, $\log(\cwc)=0.6673(5).$

We also studied the ISAW series at $\cw=\cwc,$ by the method of differential approximants. Just as in the non-interacting case discussed above, we found two singularities. The dominant singularity at $\ew=\ew_c \approx 0.309805,$ with an exponent of about $1.12,$ compared to the predicted value of $\frac87=1.1428\ldots,$ and a second singularity at $\ew=-\ew_c$ with a small exponent that is difficult to estimate, but appears to be about $-0.1 \pm 0.05.$ Both $u_c$ and $\ew_c$ were estimated in earlier work by Grassberger and Hegger \cite{GH95}, who estimated $u_c = 0.665 \pm 0.002$ and $\ew_c=0.3101 \pm 0.0004,$ both of which are in agreement with our more precise estimates.

\subsection{Square lattice bridges and TAWs at the theta point}\label{ssec:bridges_taws}

We have also generated rather short series for interacting bridges and interacting terminally attached self-avoiding walks (TAWs) on the square lattice, by a simple backtracking algorithm. This allows us to give moderately accurate estimates of the exponents at the $\theta$ point for these quantities. For TAWs an earlier estimate is given in \cite{FOT92}, but as far as we are aware, no previous estimate for interacting bridges has been given.

In \cite{SS88} the estimate $\gamma_1(\theta) = 0.57 \pm 0.09$ was given, based on a Monte Carlo analysis, using their earlier estimate of the $\theta$ point, $\cwc = 1.915 \pm 0.06$. In \cite{FOT92} the estimate $\gamma_1(\theta) = 0.57 \pm 0.02$ was given, based on a 28 term series, from which the $\theta$ point was estimated to be at $\cwc = 1.93 \pm 0.03$. Our analysis, based on a 30 term series, using differential approximants, and assuming our estimate $\cwc=1.9474$ gives a slightly lower estimate, $\gamma_1(\theta) = 0.55 \pm 0.03,$ but all three calculations of this exponent are essentially in agreement. Note that this exponent was predicted to be exactly $\frac47=0.57142\ldots$ by Duplantier and Saleur \cite{DS87}, albeit for a modified version of the interacting SAW model, being a loop model with annealed vacancies. The numerical results would appear to be in agreement with this prediction.

More precisely, Duplantier and Saleur \cite{DS87}  first predicted the boundary exponents $\gamma_1=8/7$ and $\gamma_{11}=4/7.$  However, these were exponents characteristic of the special transition, not of the ordinary transition. This "shift", until it was recognised, led to a lot of discussion in the literature about the possible differences between the standard $\theta$ point  and so-called $\theta'$ point of the Duplantier-Saleur vacancy percolation model. Actually, the $\theta$ and $\theta'$ fixed points are the same. 

In \cite{VSS91} Vanderzande et al.\ gave a persuasive argument, using connections with site percolation on the two-dimensional triangular lattice, for the resolution of the puzzle of the exponents at the ordinary and special transitions. They also used series analysis to estimate $\gamma_1 = 1.11\pm0.04$ for the special transition and $0.57\pm0.02$ for the ordinary transition. (See also \cite{FOT92} for later work.) Finally the value of $\gamma_1(\theta)=4/7$ in the ordinary case was correctly calculated by the same authors in \cite{SSV93} from the Duplantier-Saleur exponents for three lines attached to the boundary (not one) via a string of geometrical arguments. The reason that the Duplantier-Saleur exponents are those of the special transition is the fact that the ISAW at the $\theta$-point touches and bounces on the boundary many times when the transition is driven by vacancies (as does $SLE_6$ for percolation). 

For bridges at the $\theta$ point, we estimate $\gamma_b(\theta) = 0.00 \pm 0.03,$ again based on a differential approximant analysis. As far as we are aware, this exponent has not previously been estimated. In \cite{DG19b} this exponent is calculated for the first time, along with a number of other exponents for confined polymer networks, and is predicted to be precisely zero.

\section{Conclusion}\label{sec:conclusion}

For the honeycomb lattice we estimate $\cwc = 2.767 \pm 0.002.$ The honeycomb lattice is unique among the regular two-dimensional lattices in that the exact growth constant is known for non-interacting walks. It is $\sqrt{2+\sqrt{2}}$ \cite{DS12}, while for half-plane walks interacting with a surface, the critical fugacity, again for the honeycomb lattice, is $1+\sqrt{2}$ \cite{BBDDG}. We could not help but notice that $\sqrt{2+4\sqrt{2}} = 2.767\ldots .$ We have been unsuccessful in trying to devise a proof strategy, analogous to that used to prove the two known results just quoted, to try and prove this result. At present it remains a possibility, but it would be wrong to even dignify it with the title `conjecture'.
For square lattice bridges and TAWs at the $\theta$-point we estimate the exponents to be $\gamma_b=0.00 \pm 0.03,$ and $\gamma_1=0.55 \pm 0.03$ respectively. The latter result is consistent with the prediction \cite{DS87, SS88, SSV93} $\gamma_1(\theta) = \nu =\frac47.$
\section*{Acknowledgements}

We thank Bertrand Duplantier and Gordon Slade for helpful comments on an earlier version of this paper. NRB was supported by the Australian Research Council grant DE170100186. AJG acknowledges support from ACEMS. The computations for this project were undertaken with the assistance of resources and services from the National Computational Infrastructure (NCI), which is supported by the Australian Government.

\end{document}